# The subtlety of the outermost stellarator magnetic surface

Alkesh Punjabi
Department of Mathematics, Hampton University, Hampton, VA 23668, USA
and
Allen H. Boozer
Department of Applied Mathematics and Applied Physics, Columbia University, New York, NY 10027, USA

**Abstract:**
An analytic model of the magnetic field line behavior in a stellarator is used to study the subtlety of the concept of an outermost magnetic surface.  The analytic model that we use has a central region of nested magnetic surfaces.  The outermost perfectly confining surface has a toroidal flux of 0.86 of the toroidal flux of the outermost confining surface.  The field lines in the annulus between these surfaces strike a distant wall, but they make tens of thousands of transits through a period of the stellarator before doing so.  The number of transits is so large that this region can probably be viewed as having confining surfaces.  Between the outermost confining surface and a surface at 1.02 times the toroidal flux field lines go to the walls in four flux tubes: two with inward flux and two with outward flux.  One of the inward-outward pairs of flux tubes are adjoining and the other pair is separated.  When the toroidal flux is greater than 1.02 of that of the outermost confining surface approximately 85% of the field lines strike the wall before transiting a single period.  The loss of plasma from this region is so fast compared to cross-field plasma diffusion that they are probably irrelevant to the study of divertors. In addition to the two inward-outward flux tube pairs that escape from the region inside 1.02 times the flux of the confining region, the outer region has two new inward-outward of flux-tube pairs; one adjoining and one separated.

## I. Introduction

Basically, there are two types of divertors for stellarators: the island divertor and the nonresonant divertor. The island divertor is extensively studied [1-7]. It is used in the W7-X stellarator [8-12]. The nonresonant divertor is a potential alternative to the island divertor for stellarators [13-15]. The nonresonant divertor is relatively understudied. The nonresonant divertor concept needs to be thoroughly investigated to evaluate its potential as a divertor solution for stellarator fusion power plants.

There are two ways to study the nonresonant divertor: To use the magnetic configuration of actual devices such as HSX [16] or CTX [17] as in [18-22] or use an analytic model of nonresonant divertor as in [23-25**]**. An analytic model has an explicit mathematical expression for the Hamiltonian of the field lines in canonical coordinates. The Hamiltonian for the field lines is the poloidal magnetic flux as a function of the toroidal flux, a poloidal angle, and a toroidal angle. In this paper, we use the analytic model to study the subtleties of the concept of an outermost confining magnetic surface.

In the analytic model, the outermost surface is at $(r/b, \theta, \zeta) = (0.87, 0, 0)$. $r/b$ is radial position, $\theta$ is the poloidal angle, and $\zeta$ is the toroidal angle of a single period. $\zeta$ is related to the toroidal angle of the stellarator $\varphi$ by $\zeta = n_p \varphi$. $n_p$ is the number of periods of the stellarator. The wall is a circular torus with radius $r_{WALL}/b = 4$. $b$ is a nominal minor radius. The physical meaning of the parameter $b$ is hard to express. However, all the fractions of $b$ have been converted by taking squares into fractions of the toroidal flux of the outermost confining surface, which is a common way to express things. The radii of surfaces are referred to the radial positions $r/b$ of the surfaces when the poloidal angle $\theta = 0$ in the poloidal plane $\zeta = 0$ with $r = \sqrt{\psi_t}$ where $\psi_t$ is the toroidal flux of a surface.

In our previous studies of the nonresonant divertor [23-25], the analytic Hamiltonian for field lines in stellarators was derived [23], and used to calculate the trajectories of field lines in nonresonant stellarator divertor. In these papers, these trajectories represented the lines that start just outside the outermost confining surface $r/b = 0.87$ in the poloidal plane $\zeta = 0$. The lines were started in a thin annulus of width $\Delta r/b = 0.01$ surrounding the outermost surface $r/b = 0.87$ in the poloidal plane $\zeta = 0$. The points where the field lines struck the wall were the magnetic footprints which had the shape of helical stripes. The lines were integrated both in the forward direction (meaning increasing toroidal angle) and the backward direction (meaning decreasing toroidal angle). The forward moving lines are the outgoing lines, and the backward moving lines are the incoming lines. The forward/outgoing lines go from the annulus outside the outermost surface to the wall, and the backward/incoming lines go from the wall to the annulus. It was found that both the outgoing and incoming lines traveled in pairs of magnetic flux tubes – outgoing lines traveled to the wall in one tube and the incoming lines traveled in the other tube. The magnetic flux tubes always come in a pair - an outgoing tube and an incoming tube. The magnetic fluxes in the outgoing and incoming tubes of each pair were equal. This is because the magnetic field is divergence-free, $\nabla \cdot \boldsymbol{B} = 0$. One of the pairs was an adjoining pair and the other was a separated pair. In the adjoining pair, the outgoing tube and the incoming tube start at adjacent locations outside the outermost surface. In the separated pair, the outgoing tube and the incoming tube start at separate locations outside the outermost surface [24]. A method for the calculation of 3D structure of the magnetic flux tubes was developed for these field lines [25].

In this paper, we expand the horizon of our exploration of nonresonant divertor. For this purpose, we divide the annulus bound by the outermost perfectly confining magnetic surface $r/b = 0.81$ and the wall $r_{WALL}/b = 4$ in the poloidal plane $\zeta = 0$ into three annuli. The first annulus is bound on the inside by the outermost perfectly confining surface $r/b = 0.81$ and the outermost surface $r/b = 0.87$. The first annulus is populated by confining surfaces. The toroidal flux of the first annulus has 0.86 of the toroidal flux of the outermost surface. The second annulus surrounds the outermost surface. The thickness of the second annulus is $\Delta r/b = 0.01$. The toroidal flux of the second annulus is $\cong 0.02$ of the toroidal flux of the outermost surface.

The second annulus was studied in our previous papers [23-25]. The third annulus is bound on the inside by $r/b = 0.88$ and the wall $r_{WALL}/b = 4$. The third annulus has toroidal flux which is greater than 1.02 of the toroidal flux of the outermost surface. The radial positions here are when $\theta = 0$ in the poloidal plane $\zeta = 0$. The three sub-annuli are ordered by how far out they are started from the magnetic axis. We also refer to each of the three sub-annuli as a region. Figure 1 shows the three sub-annuli/regions.

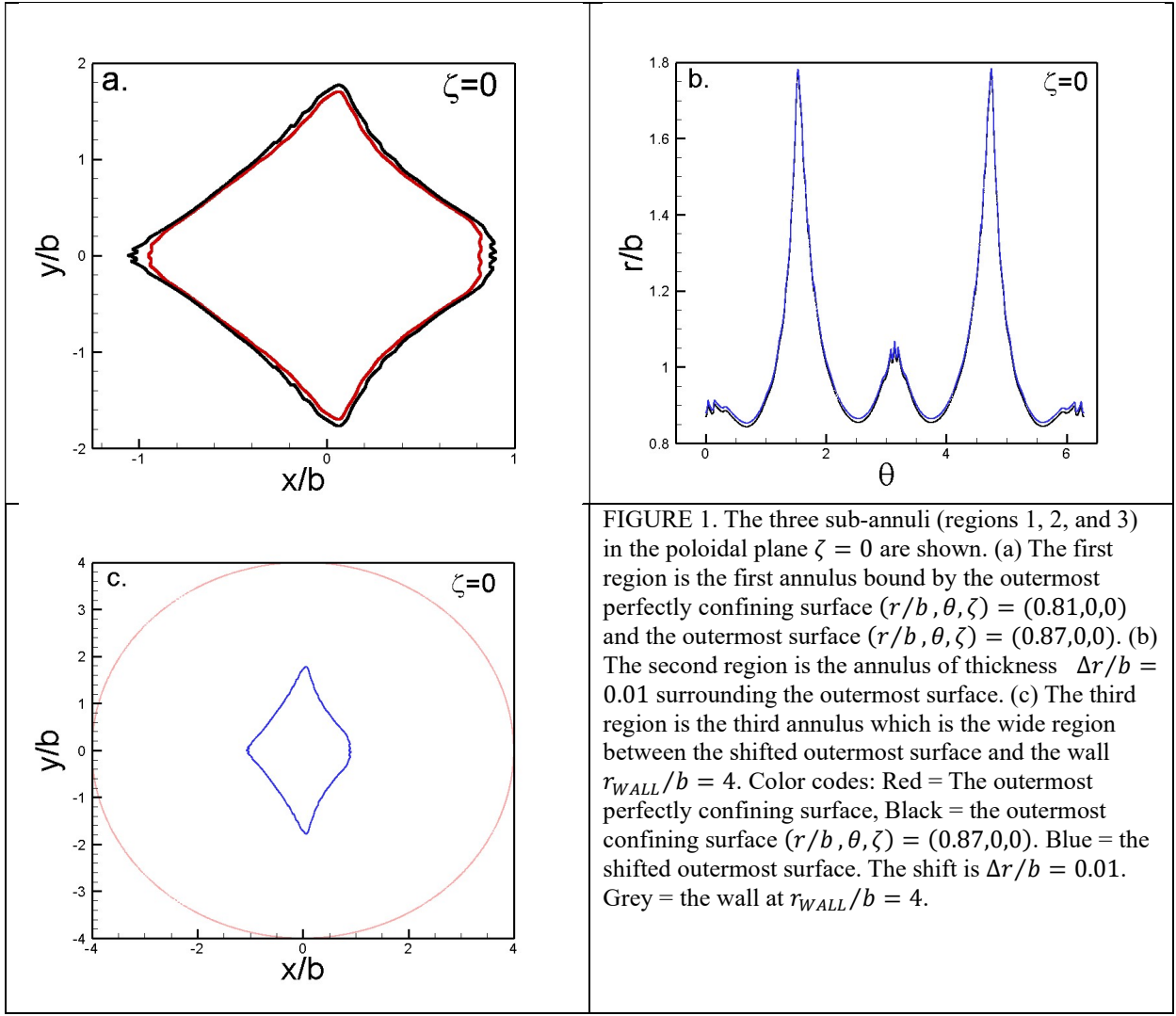

FIGURE 1. The three sub-annuli (regions 1, 2, and 3) in the poloidal plane $\zeta = 0$ are shown. (a) The first region is the first annulus bound by the outermost perfectly confining surface $(r/b, \theta, \zeta) = (0.81,0,0)$ and the outermost surface $(r/b, \theta, \zeta) = (0.87,0,0)$. (b) The second region is the annulus of thickness $\Delta r/b = 0.01$ surrounding the outermost surface. (c) The third region is the third annulus which is the wide region between the shifted outermost surface and the wall $r_{WALL}/b = 4$. Color codes: Red = The outermost perfectly confining surface, Black = the outermost confining surface $(r/b, \theta, \zeta) = (0.87,0,0)$. Blue = the shifted outermost surface. The shift is $\Delta r/b = 0.01$. Grey = the wall at $r_{WALL}/b = 4$.

In the first region, field lines are started in the thin annuli of width $\Delta r/b = 0.01$ surrounding the confining surfaces $r/b = 0.81, 0.82, 0.83,$ and $0.84$. All these surfaces are inside the outermost surface. It is found that surfaces $r/b = 0.82, 0.83,$ and $0.84$ are leaky. This means that a small fraction of lines is able to leak through these surfaces after hundreds of thousands of toroidal transits of a period and then strike the wall. The segments of these surfaces where there are sharp edges and changes in curvature are found to be sticky. On these segments the field lines linger for long times. These segments are identified. The surfaces in this region probably act as confining surfaces, and the $r/b = 0.81$ is found to be the outermost perfectly confining surface. The maximum loss is 5.3%. For the outermost perfectly confining surface the loss is zero.

The second region is the thin annulus of width $\Delta r/b = 0.01$ surrounding the outermost surface $r/b = 0.87$. The behaviour of field lines in this annulus is described above and in detail in [24-25].

In the third region, a uniform $(\theta, r/b)$ grid is formed, and field lines are started on the grid points and advanced in the forward and backward direction for 10K toroidal circuits of the period. The field lines in this region very quickly strike the wall in either direction, FW or BW. They typically make less than one toroidal transit before striking the wall. A few percent of the lines interact with the thin second annulus. Lines from this third annulus generate two new pairs of flux tubes – one adjoining pair and one separated pair. The angular widths of the tubes become narrower with increasing radial distance. The phase portraits of lines indicate the presence of four X-lines.

The points of interception of the field lines with the wall $r_{WALL}/b = 4$ give the magnetic footprint. Magnetic footprints are helical stripes. The field lines are not stopped when they intercept the wall. They are continued until their radial position crosses $r/b = 10$. This is done to get larger picture of the magnetic flux tubes.

This paper is organized as follows: Section II gives the model analytic Hamiltonian and the map equations for the field lines. Sections III, IV, and V describe the behavior of the field lines in the first, second, and the third sub-annuli or regions, respectively. Section VI gives the conclusions and the discussion of results.

## II. The model Hamiltonian

An analytic model means an explicit expression for the Hamiltonian of the magnetic field lines in canonical coordinates. This is the poloidal magnetic flux as a function of the toroidal magnetic flux, a poloidal angle, and a toroidal angle. The vanishing of the divergence of the magnetic field together with then non-vanishing of the magnetic field strength in the region of interest guarantees such a function exists, and it must be analytic when the magnetic field is an analytic function of position. The analytic model that we use has a central region of nested magnetic surfaces. The analytic Hamiltonian for the trajectories of magnetic field lines in the nonresonant divertor was derived in [23]. The Hamiltonian is the poloidal flux $\psi_p(\psi_t, \theta, \zeta)$. $\psi_t, \theta,$ and $\zeta$ are canonical coordinates. $\psi_t$ and $\theta$ are conjugate for evolution in $\zeta$ with $\theta$ the canonical position and $\psi_t$ the canonical momentum. The analytic Hamiltonian is

$$\psi_p(\psi_t, \theta, \zeta) = \left[\iota_0 + \frac{\varepsilon_0}{4}\left((2\iota_0 - 1)cos(2\theta - \zeta) + 2\iota_0 cos(2\theta)\right)\right]\psi_t$$
$$+ \frac{\varepsilon_t}{6}[(3\iota_0 - 1)cos(3\theta - \zeta) - 3\iota_0 cos(3\theta)]\psi_t^{3/2}$$
$$+ \frac{\varepsilon_x}{8}[(4\iota_0 - 1)cos(4\theta - \zeta) + 4\iota_0 cos(4\theta)]\psi_t^2. \qquad (1)$$

$\psi_p$ is the normalized poloidal flux, $\psi_t$ is the normalized toroidal flux, $\zeta$ is the toroidal angle per period, and $\iota_0$ is the rotational transform on the magnetic axis. The parameters $\varepsilon_0$, $\varepsilon_t$, and $\varepsilon_x$ control the shape of magnetic surfaces. They are called shape-parameters. $\varepsilon_0$ controls the elongation of magnetic surfaces, $\varepsilon_t$ controls the triangularity, and $\varepsilon_x$ controls the sharpness of the edges on the confining surfaces. $\iota_0$ is the rotational transform on the magnetic axis.

The map equations for the field line trajectories are [24]

$$\psi_t^{(j+1)} = \psi_t^{(j)} - \frac{\partial \psi_p\left(\psi_t^{(j+1)}, \theta^{(j)}, \zeta^{(j)}\right)}{\partial \theta^{(j)}} \delta\zeta; \quad (2)$$

$$\theta^{(j+1)} = \theta^{(j)} + \frac{\partial \psi_p\left(\psi_t^{(j+1)}, \theta^{(j)}, \zeta^{(j)}\right)}{\partial \psi_t^{(j+1)}} \delta\zeta; \quad (3)$$

$$\zeta^{(j+1)} = \zeta^{(j)} + \delta\zeta. \quad (4)$$

$j$ denotes the iteration number. The step-size of the map is $\delta\zeta = 2\pi/3600$ The map preserves the symplectic invariant:

$$\frac{\partial\left(\psi_t^{(j+1)}, \theta^{(j+1)}\right)}{\partial\left(\psi_t^{(j)}, \theta^{(j)}\right)} = +1. \quad (5)$$

The shape parameters that were used are: $\varepsilon_0 = 1/2$, $\varepsilon_t = 1/2$, and $\varepsilon_x = -0.31$. The rotational transform is $\iota_0 = 0.15$. The outermost surface is at $(r/b, \theta, \zeta) = (0.87, 0, 0)$.

Fractions of $b$ are converted to the fractions of toroidal magnetic flux of the outermost surface using $r = \sqrt{\psi_t}$. The radius of a surface is given in terms of the radial position of the surface when the poloidal angle $\theta = 0$ in the poloidal plane $\zeta = 0$. The wall is at $r_{WALL}/b = 4$. It is an axisymmetric circular torus. Trajectories are terminated when $r/b \geq 10$. The surface $r_T/b = 10$ is the termination surface. This surface is also a circular axisymmetric torus. The subscript $T$ denotes termination surface. The stellarator has five periods, $n_P = 5$. All the parameters are same as used in [23-25].

### III. The first region: Leaky and sticky surfaces

The first annulus is bound by the surfaces $r/b = 0.81$ and the outermost surface $r/b = 0.87$. In the first annulus, four surfaces, $r/b = 0.81, 0.82, 0.83$, and $0.84$, are chosen. We call these surfaces starting surfaces. A thousand points on a given starting surface are chosen in the poloidal plane $\zeta = 0$. The radial position of each point on a given starting surface is shifted radially outward through a maximum distance of $\Delta r/b = 0.01$. The shift is random meaning that the shift $\delta r/b = R_N(\Delta r/b)$. $R_N$ is a random number in the interval $(0,1]$. These one thousand lines for a given starting surface are advanced for 200 K transits of a period in FW direction. See Table 1. A small fraction of lines, maximum of about 5%, leak through the intervening surfaces and strike the wall. See Figure 2. The leakage becomes smaller as the starting surface moves away from the outermost surface $r/b = 0.87$. No leakage occurs for lines starting within $\Delta r/b \leq 0.01$ of the starting surface $r/b = 0.81$. See Table 1.

Table 1. The starting surface and the number of lines leaking to the wall

| Starting surface | # of lines that leak to the wall |
|---|---|
| $r/b = 0.84, \theta = 0, \zeta = 0$ | 53 |
| $r/b = 0.83, \theta = 0, \zeta = 0$ | 17 |
| $r/b = 0.82, \theta = 0, \zeta = 0$ | 2 |
| $r/b = 0.81, \theta = 0, \zeta = 0$ | 0 |

In the Figure 2, we show the Poincaré plot of the field line trajectories for the starting surfaces $r/b = 0.84, 0.83$, and $0.82$ in the $\zeta = 0$ poloidal plane. The figure shows that the leaking field lines go into two outgoing tubes. The tube on the right half is the outgoing tube of the adjoining pair, and the tube in the bottom half is the outgoing tube of the separated pair [25].

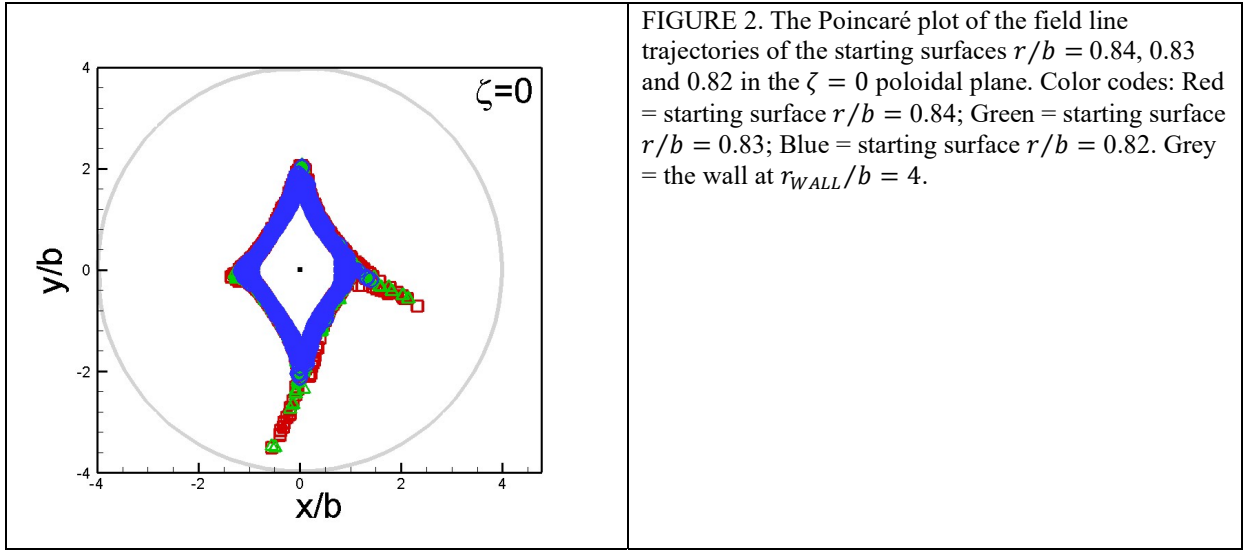

FIGURE 2. The Poincaré plot of the field line trajectories of the starting surfaces $r/b = 0.84$, $0.83$ and $0.82$ in the $\zeta = 0$ poloidal plane. Color codes: Red = starting surface $r/b = 0.84$; Green = starting surface $r/b = 0.83$; Blue = starting surface $r/b = 0.82$. Grey = the wall at $r_{WALL}/b = 4$.

The leak-time, $\tau_{leak}$, is the number of toroidal transits of a period it takes for a field line to leak and strike the wall, $\tau_{leak} = \zeta_{strike}/2\pi$. The lines start in the poloidal plane $\zeta = 0$. $\zeta_{strike}$ is the toroidal angle covered by a line when it strikes the wall. In Figure 3, we show the leak times for starting surfaces $r/b = 0.84, 0.83$, and $0.82$. The leak-time can be as small as two hundred transits and as long as hundreds of thousands of transits. See Figure 3. For the starting surface $r/b = 0.84$, the number of lines that have not yet leaked drops exponentially. This means that the loss of lines for the starting surface $r/b = 0.84$ is random. For the starting surfaces $r/b = 0.83$ and $0.82$, the drop is not exponential. This looks as if there is a special region in the annulus ($0.82 \leq r/b \leq 0.84$) from which lines are lost with otherwise far better confinement.

In Figure 4, we show the footprint of the field lines which leak and strike the wall. In Figure 4, we also show the footprint of the field lines for the starting surface $r/b = 0.87$ which is the outermost surface. In Figure 4, the number of FW lines that strike the wall for the outermost surface is 645 (out of 1000 lines) [25]. The upper stripe in Figure 4 is the outgoing tube of the adjoining pair and the lower stripe is the outgoing tube of the separated pair [25]. The Figure 4 is as expected. Note that the blue points ($r/b = 0.82$) must abut the green ($r/b = 0.83$) and the green the red ($r/b = 0.84$). The points from inner annuli must pass through the outer annuli as well.

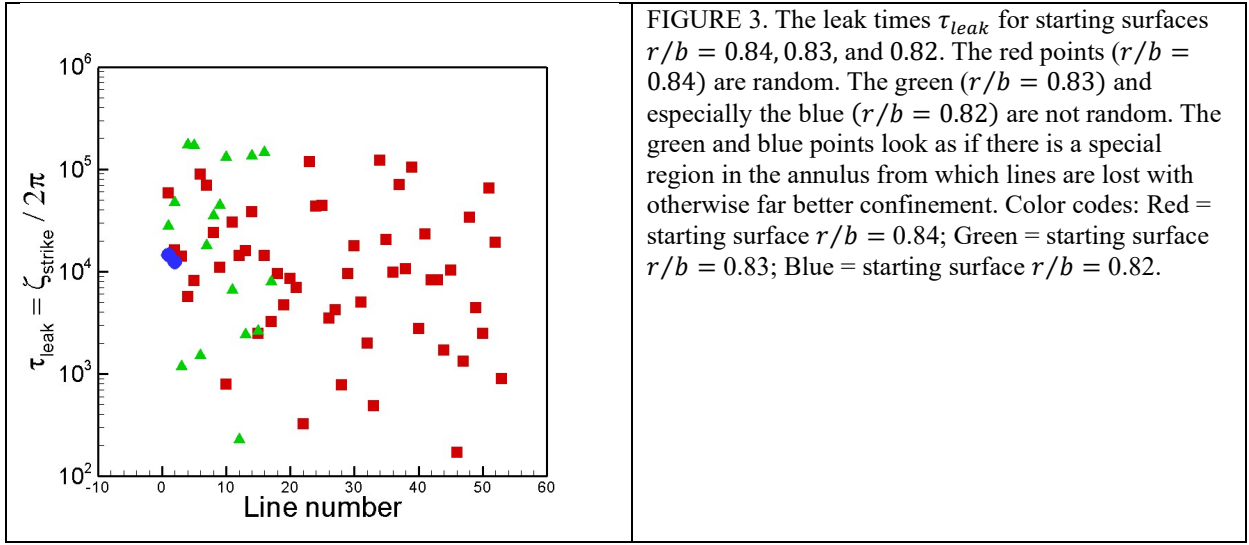

FIGURE 3. The leak times $\tau_{leak}$ for starting surfaces $r/b = 0.84, 0.83,$ and $0.82$. The red points ($r/b = 0.84$) are random. The green ($r/b = 0.83$) and especially the blue ($r/b = 0.82$) are not random. The green and blue points look as if there is a special region in the annulus from which lines are lost with otherwise far better confinement. Color codes: Red = starting surface $r/b = 0.84$; Green = starting surface $r/b = 0.83$; Blue = starting surface $r/b = 0.82$.

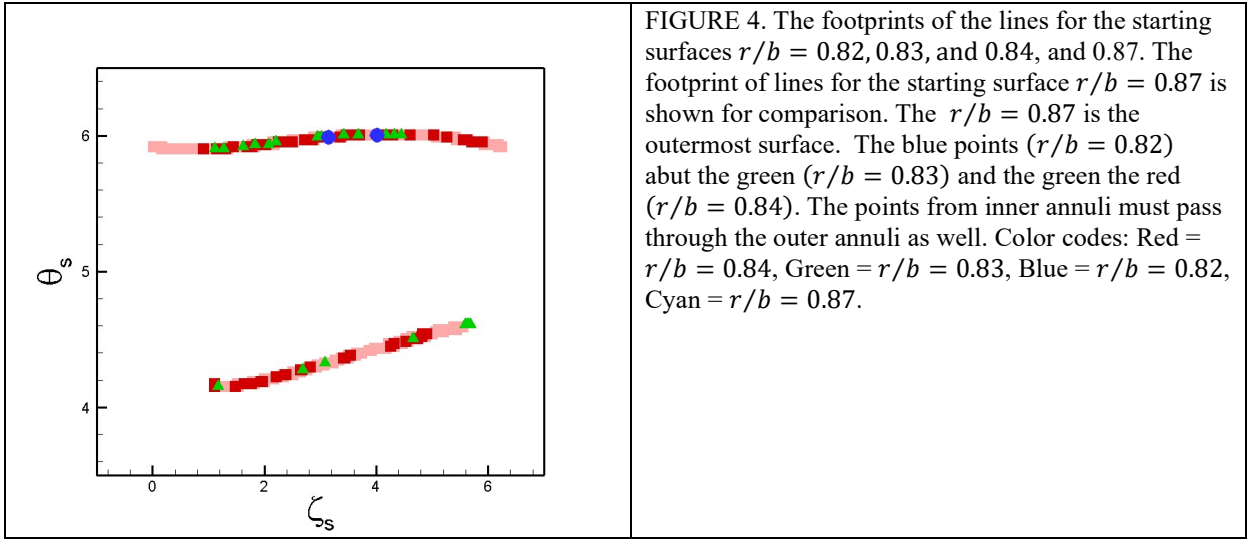

FIGURE 4. The footprints of the lines for the starting surfaces $r/b = 0.82, 0.83,$ and $0.84,$ and $0.87$. The footprint of lines for the starting surface $r/b = 0.87$ is shown for comparison. The $r/b = 0.87$ is the outermost surface. The blue points ($r/b = 0.82$) abut the green ($r/b = 0.83$) and the green the red ($r/b = 0.84$). The points from inner annuli must pass through the outer annuli as well. Color codes: Red = $r/b = 0.84$, Green = $r/b = 0.83$, Blue = $r/b = 0.82$, Cyan = $r/b = 0.87$.

In the poloidal plane $\zeta = 0$, the circular area bounded by the wall $r_{WALL} = 4b$ is divided in $360 \times 400$ cells of equal size $\Delta\theta\Delta(r/b) = (2\pi/360) \times (b/100)$ in the $(\theta, r/b)$ plane. Each cell is labelled by two indices, $i$ and $j$, $1 \leq i \leq 360$, $\leq 1 \leq j \leq 400$. The angular position of a cell is given by the index $i$, $(i-1)(2\pi/360) < \theta \leq i(2\pi/360)$. The radial position of a cell is given by the index $j$, $(j-1)(b/100) < r/b \leq j(/100)$. Each field line is advanced for 200 K transits. Occupation count of a cell is the number of crossings of that cell by the trajectories of field lines. It is denoted by $N_{ij}$. It is initialized to zero, and every time a line crosses the cell, it is raised by unity. If no field line ever crosses a cell, the occupation count of that cell is zero, $N_{ij} = 0$. If the occupation count of a cell is non-zero, $N_{ij} \geq 1$, the cell is called an occupied cell. The average occupation count for the occupied cells is denoted by $\langle N_{occupied}\rangle$. A cell is called "sticky" if $1 \leq N_{ij}/\langle N_{occupied}\rangle \leq 5$. A cell is called "most-sticky" if $N_{ij}/\langle N_{occupied}\rangle > 5$. A cell is "non-sticky" if $N_{ij}/\langle N_{occupied}\rangle < 1$. This classification helps to identify the stickiness of the cells in the first annulus.

In Table 2, we give the percentage of cells that are sticky and most-sticky and the average occupation count for the starting surfaces $r/b = 0.81, 0.82, 0.83,$ and $0.84$. From Table 2, we see that as we move radially out and approach the outermost surface, the surfaces become stickier and the average occupation count becomes smaller.

Table 2. Percentage of cells that are sticky and most-sticky
and the average occupation count for starting surfaces

| Surface | Most-sticky (%) | Sticky (%) | Average $\langle N_{occupied}\rangle$ |
|---|---|---|---|
| $r/b = 0.81$ | 0.81 | 37.00 | 1077005 |
| $r/b = 0.82$ | 3.66 | 25.50 | 745432 |
| $r/b = 0.83$ | 4.35 | 22.35 | 618486 |
| $r/b = 0.84$ | 5.99 | 17.97 | 516576 |

In Figure 5, we show the sticky cells. Figure 5 shows that the cells that are located on and close to the starting surfaces are sticky. Cells close to and outside the surfaces are also sticky. In Figure 6, we show the most-sticky cells. In Figure 6, we see that the most-sticky cells are located at specific locations on and close to the starting surfaces. These specific locations are shown in the Figures 6b-f. Figures 6a-f tells us that the most-sticky cells are located where the surfaces have sharp edges and where the curvature of surfaces change.

No field line leaks for the starting surface $r = 0.81b$. Therefore, $r/b = 0.81$ is the *outermost perfectly confining surface*. A thick sticky and leaky annular layer of width $w/b \sim 0.06$ exists outside this surface. Lines in this layer can take up to hundreds of thousands of transits to leak to the wall. The connection length is $L_c = 2\pi R_0 N_T/n_p$. $R_0$ is the major radius, $N_T$ is the number of transits, and $n_p$ is the number of periods. So, the lines that take hundreds of thousands of transits to strike have very long connection lengths. The surfaces inside this layer act like confining surfaces. However, a small fraction of lines can leak through these surfaces. So, these surfaces must not be perfect; they must have some openings through which lines can escape. Segments of these surfaces where there are sharp edges and changes in curvature are very sticky. This layer illustrates the behaviour of field lines in the regions of leaky and sticky surfaces which act like confining surfaces. The toroidal flux of the outermost perfectly confining surface is 0.86 of the outermost surface.

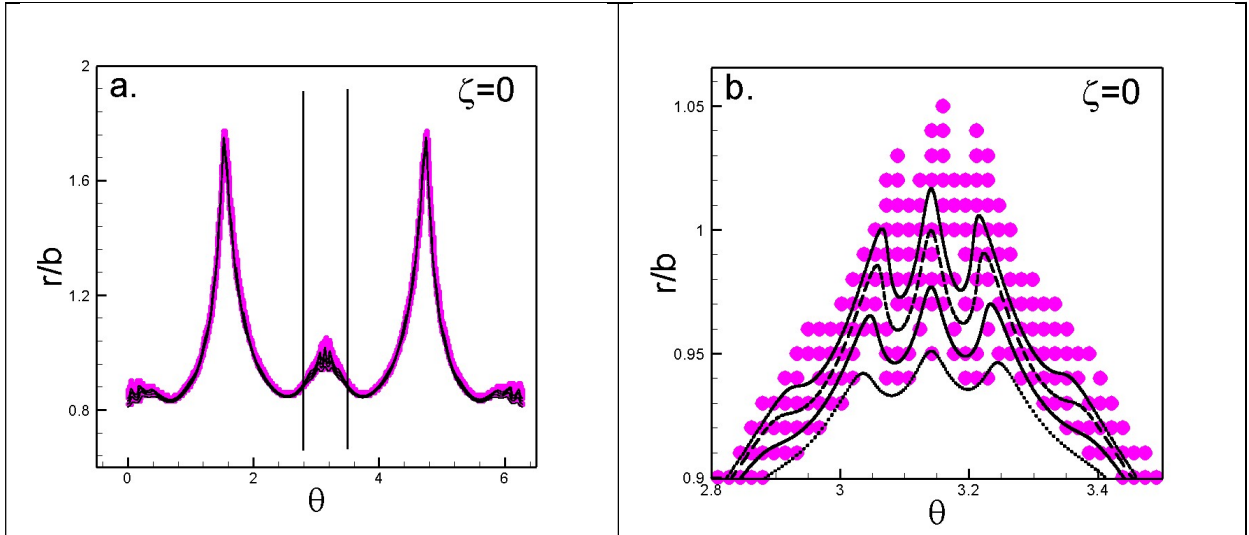

FIGURE 5. (a) The sticky cells in the in the poloidal plane $\zeta = 0$. The two vertical lines $\theta = 2.8$ and $\theta = 3.5$ show the area whose enlarged view is shown in the Figure 5(b). (b) An enlarged view of the Figure 5a. Color codes: Red = the starting surfaces at $r/b = 0.82, 0.83,$ and $0.84$, Blue = sticky cells.

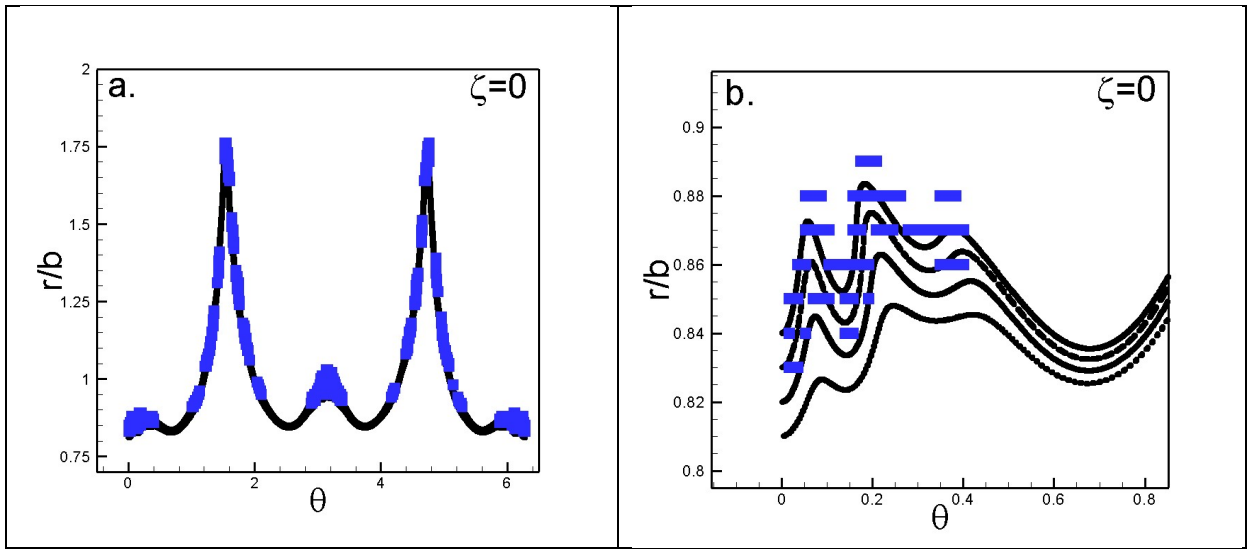

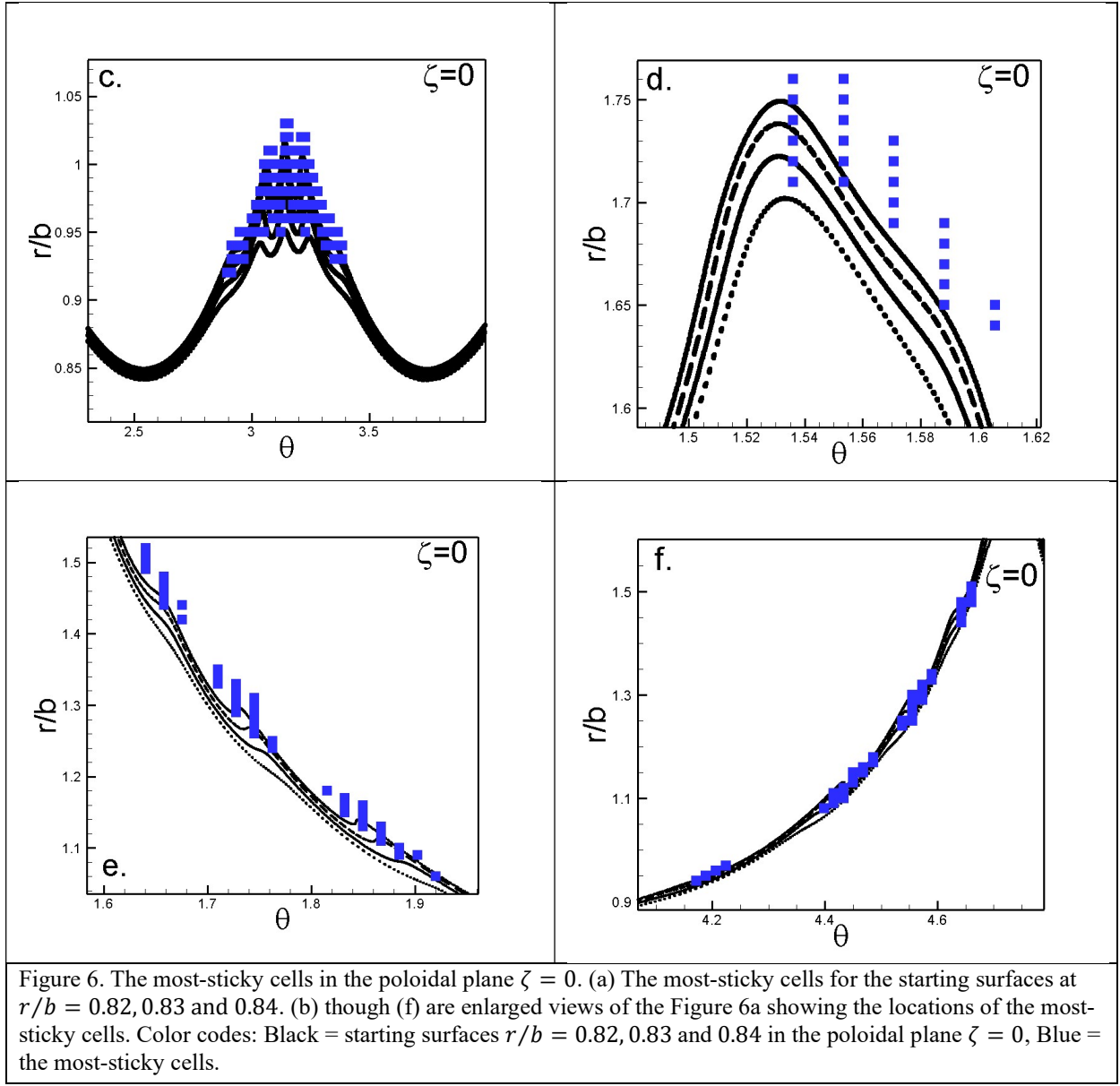

Figure 6. The most-sticky cells in the poloidal plane $\zeta = 0$. (a) The most-sticky cells for the starting surfaces at $r/b = 0.82, 0.83$ and $0.84$. (b) though (f) are enlarged views of the Figure 6a showing the locations of the most-sticky cells. Color codes: Black = starting surfaces $r/b = 0.82, 0.83$ and $0.84$ in the poloidal plane $\zeta = 0$, Blue = the most-sticky cells.

## IV. The second region: Two pairs of flux tubes

The second annulus is a narrow annulus of thickness $\Delta r/b = 0.01$ surrounding the outermost surface $r/b = 0.87$. This region is important. The plasma particles exiting the outermost surface follow the field lines that start in this annulus. The field lines starting here reach the wall through paired magnetic flux tubes.

A thousand field lines are started in this annulus and integrated FW and BW directions for 200 K transits just as done in the previous section. A close-up view of the starting positions of these lines is shown in the Figure 7. The Poincaré plots in the poloidal planes $\zeta = 0$ and $\zeta = \pi$ for the trajectories that strike the wall are shown in Figure 8. The magnetic footprints on the wall are shown in Figure 9. Figures 8 and 9 are same as in the 2022 paper [25]. They are repeated here to understand the new results in the next section. The toroidal flux inside the second annulus is about 0.02 of the outermost surface.

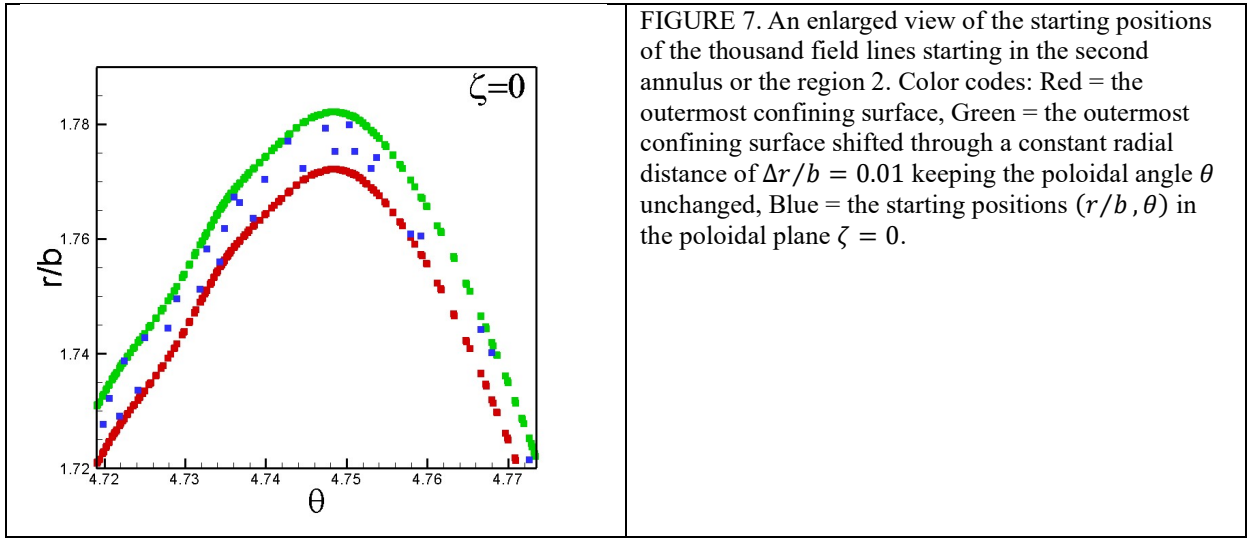

FIGURE 7. An enlarged view of the starting positions of the thousand field lines starting in the second annulus or the region 2. Color codes: Red = the outermost confining surface, Green = the outermost confining surface shifted through a constant radial distance of $\Delta r/b = 0.01$ keeping the poloidal angle $\theta$ unchanged, Blue = the starting positions $(r/b, \theta)$ in the poloidal plane $\zeta = 0$.

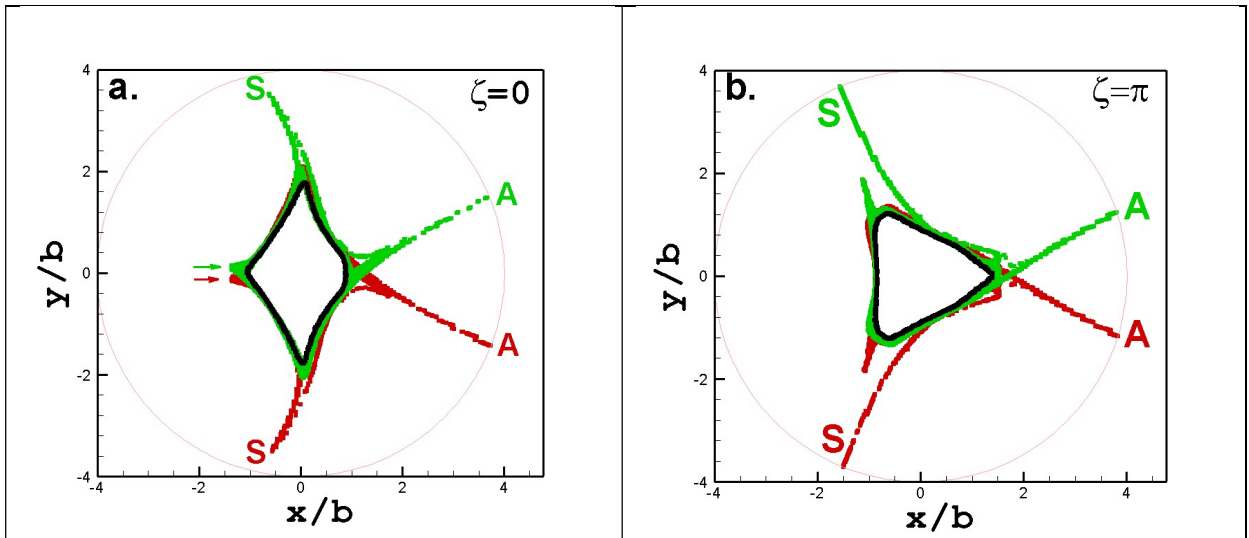

Figure 8. Poincaré plots in the poloidal planes $\zeta = 0$ and $\zeta = \pi$. The label A denotes the adjoining tubes and the label S denotes the separated tubes. The arrows indicate the pseudo-tubes. Color codes: Black = the outermost confining surface, Red = the puncture points of the FW lines which strike the wall, Green = the puncture points of BW lines that strike the wall, Grey = the wall $r_{WALL}/b = 4$.

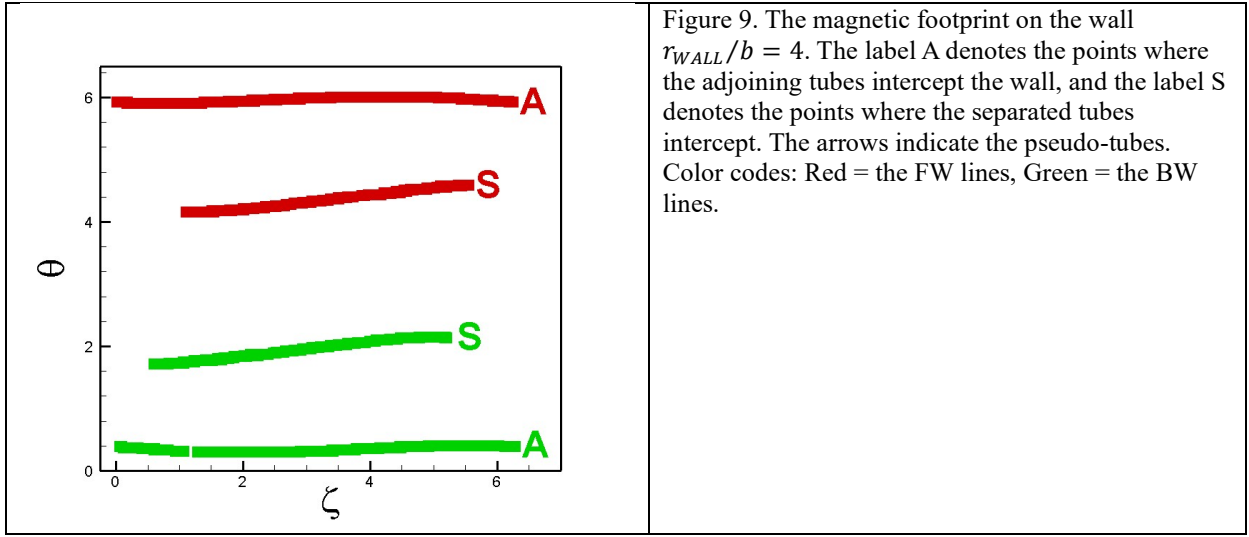

Figure 9. The magnetic footprint on the wall $r_{WALL}/b = 4$. The label A denotes the points where the adjoining tubes intercept the wall, and the label S denotes the points where the separated tubes intercept. The arrows indicate the pseudo-tubes. Color codes: Red = the FW lines, Green = the BW lines.

Figures 8 and 9 show that the forward moving lines travel from the outermost surface to the wall through two outgoing flux tubes, and the backward moving lines travel from the wall to the outermost surface through two incoming tubes. One of the outgoing tubes and one of the incoming tubes start out at adjacent locations outside the outermost surface. For this reason, these two tubes are called adjoining tubes. These two tubes form the adjoining pair. The magnetic flux in the two adjoining tubes is equal because magnetic flux is preserved. One of the two outgoing tubes and one of the incoming tubes start at separate locations outside the outermost surface. These two tubes are called the separated tubes, and the pair forms the separated pair. Again, the magnetic flux in these two separated tubes is equal. There is also a pair of pseudo-tubes where the fields go a short distance away from the outermost surface but do not reach the wall; see Figure 8. The footprint and the Poincaré plots are stellarator-symmetric [27] as expected. These results were given in [25]. The plasma particles exiting the outermost surface follow the magnetic field lines starting in the second annulus.

The loss-time of field lines in a tube, $N_T$, is given in terms of the toroidal transits of a single period. The loss-times for the tubes are shown in Figure 10.

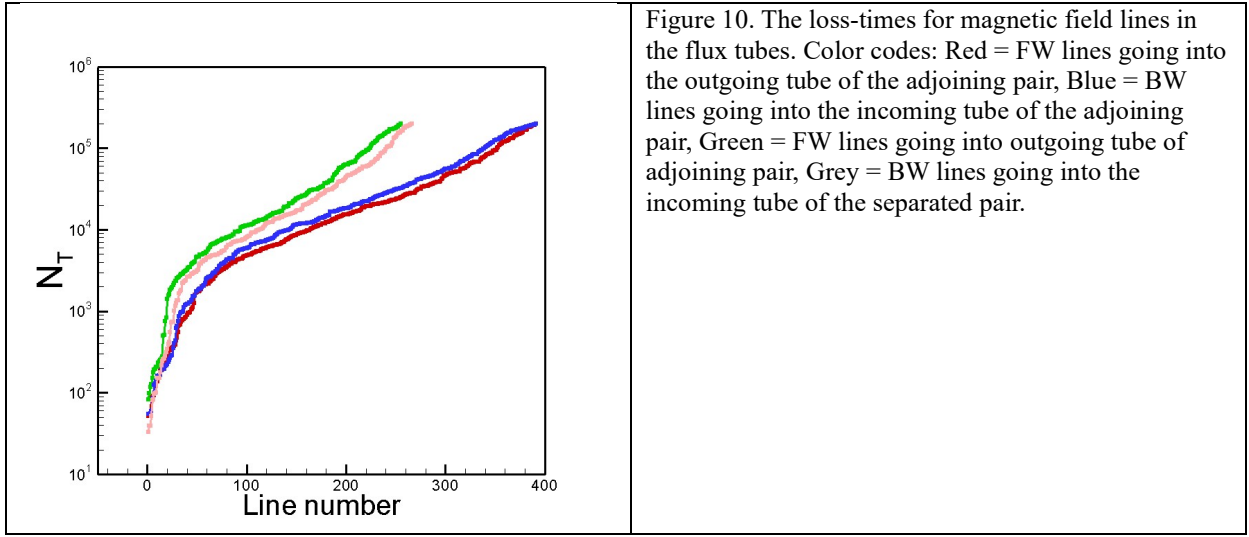

Figure 10. The loss-times for magnetic field lines in the flux tubes. Color codes: Red = FW lines going into the outgoing tube of the adjoining pair, Blue = BW lines going into the incoming tube of the adjoining pair, Green = FW lines going into outgoing tube of adjoining pair, Grey = BW lines going into the incoming tube of the separated pair.

Table 3 gives the number of transits, $N_T$, the loss-fraction, and the weighted loss-times for the FW and BW lines.

Table 3. The number of transits, $N_t$, the loss-fraction, and the weighted loss-times for the FW and BW lines.

|  | Adjoining tubes | | Separated tubes | | Loss-fraction | Weighted loss-time |
| --- | --- | --- | --- | --- | --- | --- |
|  | Lines | $N_T$ | Lines | $N_T$ | | |
| FW | 390 | 23 K | 255 | 28 K | 65 % | 25 K |
| BW | 391 | 30 K | 266 | 23 K | 66 % | 27 K |

The connection length $L_c$ is given by $L_c = 2\pi R_0 N_T / N_p$. $R_0$ is the major radius, and $N_T$ is the loss-time in transits. For W7-X, $R_0 = 5.5$ m. Then, the connection length of field lines in the W7-X will be $L_c \cong 180$ km.

It takes about 26 K transits for lines from the narrow second annulus to escape to the wall through the flux tubes. The plasma particles travelling on these lines will have ample time to diffuse across the field lines. The cross-field drifts for these particles will be large enough to be an important factor in the study of divertors. The toroidal flux of the second -annulus is about 0.02 of the toroidal flux of the outermost surface.

## V. The third region: Four pairs of flux tubes

The third annulus or region is the area between the shifted outermost surface and the. This annulus is bound on the inside by the shifted outermost surface and on the outside by the wall. A unform $(\theta, r)$ grid is formed in this area. The grid points are separated by $\Delta\theta = 2\pi/360$ and $\Delta r/b = 0.1$. The grid is $(\theta_i, r_j/b), i = 1,360$ and $j = 1,40$. This gives 10,378 grid points in the poloidal plane $\zeta = 0$. Field lines are started at these points and advanced for 10 K toroidal transits of a single period in FW and BW directions. An addition axisymmetric circular wall of radius $r_T/b = 10$ is placed outside the wall. This wall is called the termination wall. Field lines are terminated when they reach this termination wall. The starting positions of the field lines in the third annulus and the setup are shown in Figure 11.

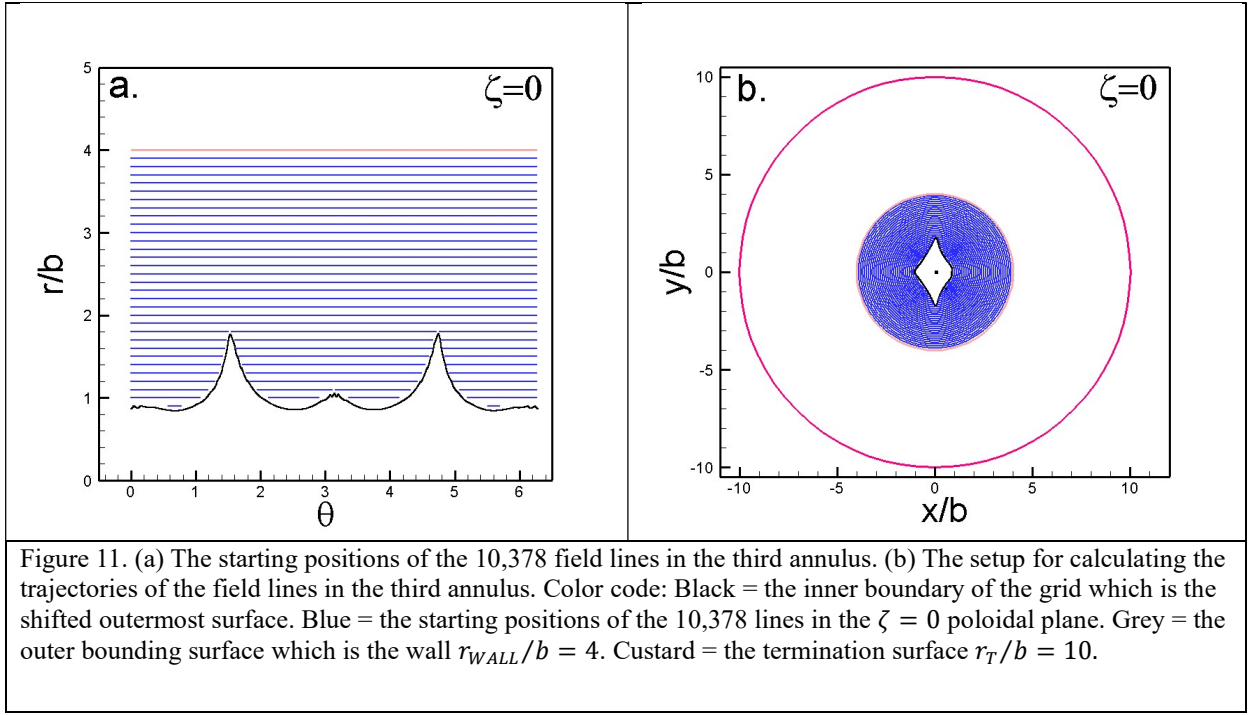

Figure 11. (a) The starting positions of the 10,378 field lines in the third annulus. (b) The setup for calculating the trajectories of the field lines in the third annulus. Color code: Black = the inner boundary of the grid which is the shifted outermost surface. Blue = the starting positions of the 10,378 lines in the $\zeta = 0$ poloidal plane. Grey = the outer bounding surface which is the wall $r_{WALL}/b = 4$. Custard = the termination surface $r_T/b = 10$.

The Poincaré plots in the poloidal planes $\zeta = 0$ and $\zeta = \pi$ are shown in Figure 12. The plots show that the magnetic field lines from the third annulus collimate and go into eight flux tubes and then strike the wall and continue inside the tubes and cross the termination surface. The plots also show that the collimation becomes sharper as the tubes move radially out. The tubes become narrower as they move out radially. In the $\zeta = 0$ plane, there appear to be five x-points. In this plane, there is one x-point on the x-axis on the right side, and two x-points on the x-axis on the left side. The two x-points on the left side appear to be x-points corresponding to the o-point of an island. There is a pair of stellarator-symmetric (up-down symmetry) x-points close to the y-axis. As we move from the poloidal plane $\zeta = 0$ to the plane $\zeta = \pi$, it appears that the island on the x-axis on the left side opens up and there are now four x-points.

*When we compare the flux tubes in Section IV with the flux tubes in this section, an important new result is found: The third annulus generates two new pairs of flux tubes, one new pair is an adjoining pair, and the other is a separated pair. In our papers in 2020 and 2022 [24,25], we found that there was one adjoining pair, a separated pair, and a pseudo-pair. The old pseudo-pair now becomes an adjoining pair. So, the important new result is that the nonresonant divertor has in all four pairs of flux tubes – two are adjoining and two are separated. See Figures 12 and 14. The footprints are again helical stripes. The toroidal flux of the third annulus is about 20 times of the outermost surface.*

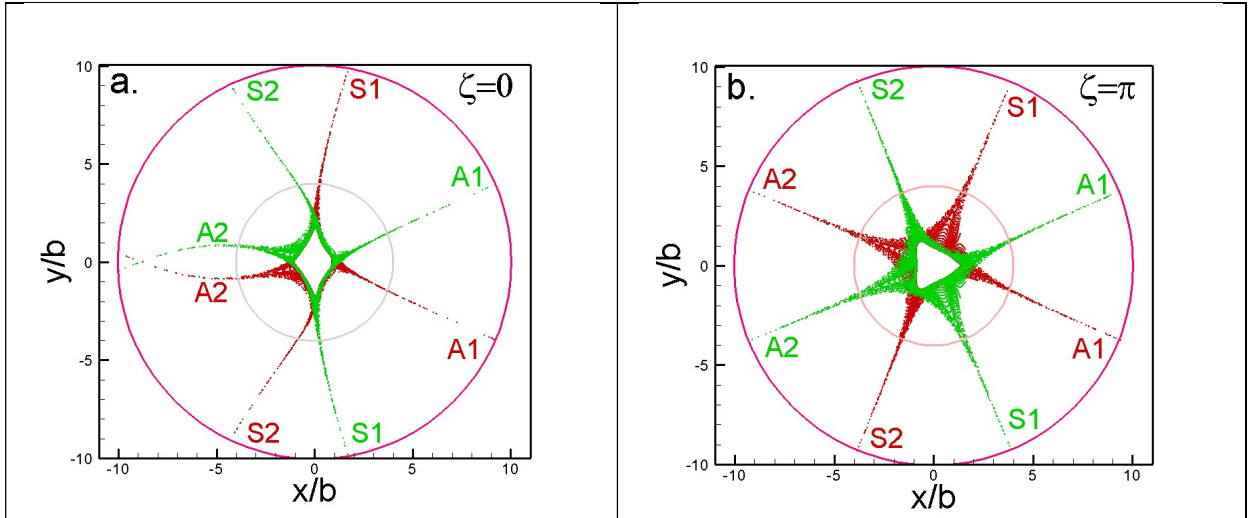

Figure 12. The Poincaré plots in the poloidal planes $\zeta = 0$ and $\zeta = \pi$ for the lines starting in the third annulus. Color codes: Red = FW trajectories, Green = BW trajectories, Grey = the wall, Custard = the termination surface. Letters A and S denote the tubes of adjoining pairs and separated pairs, respectively. Numbers 1 and 2 denote, the first and second pairs, respectively. Red letters = FW tubes, Green letters = BW tubes.

In the third annulus, the one *e*-fold loss-time is 0.46 transits of a period. 85% of field lines strike the wall within at most a single transit and the remaining 15% of lines take longer; see Figure 13. 85% of flux will strike lost before the plasma can diffuse across field lines. The footprints − are shown in Figure 14. The four outgoing tubes are shown in red and the four incoming tubes are shown in green. These eight tubes form four pairs.

In Figure 15, we show the footprints of the FW and BW lines for lines from the third annulus which take more than a single transit of a period and the footprint of lines that go into flux tubes from the second annulus. Figure 15 shows that the lines from the third annulus with more than a single transit go into the same flux tubes as lines from the second -annulus. We note that the lines from the second annulus go only into two pairs of tubes while the lines from the third annulus go into only three pairs of tubes when they make more a single transit. Lines from the third annulus that make at the most a single transit do go into the all four pairs of tubes. See Figures 14 and 15.

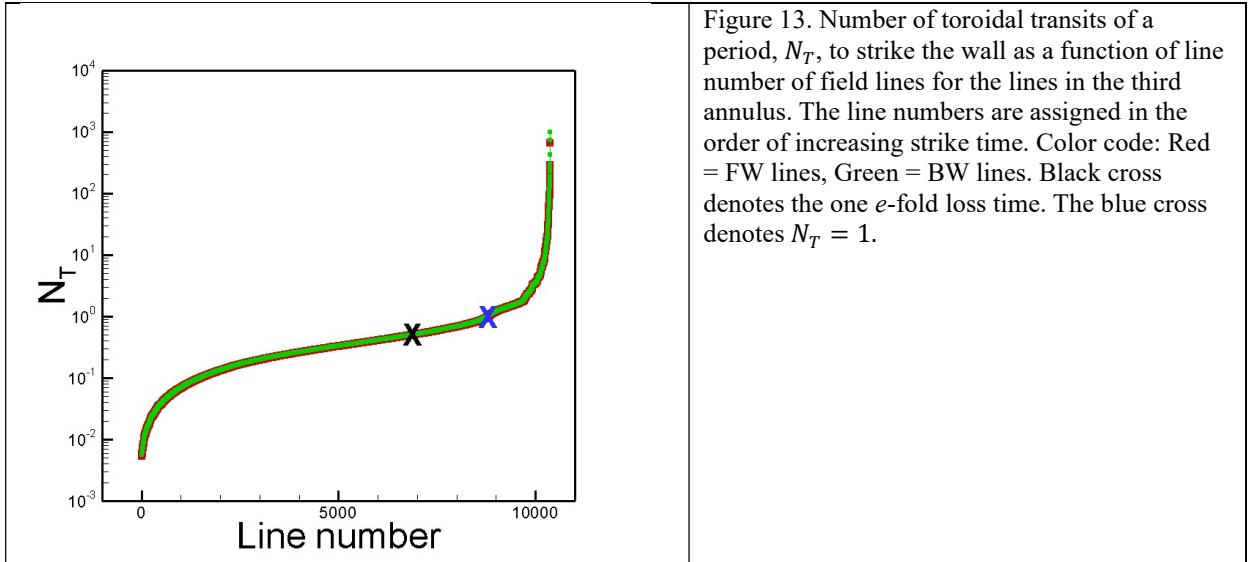

Figure 13. Number of toroidal transits of a period, $N_T$, to strike the wall as a function of line number of field lines for the lines in the third annulus. The line numbers are assigned in the order of increasing strike time. Color code: Red = FW lines, Green = BW lines. Black cross denotes the one $e$-fold loss time. The blue cross denotes $N_T = 1$.

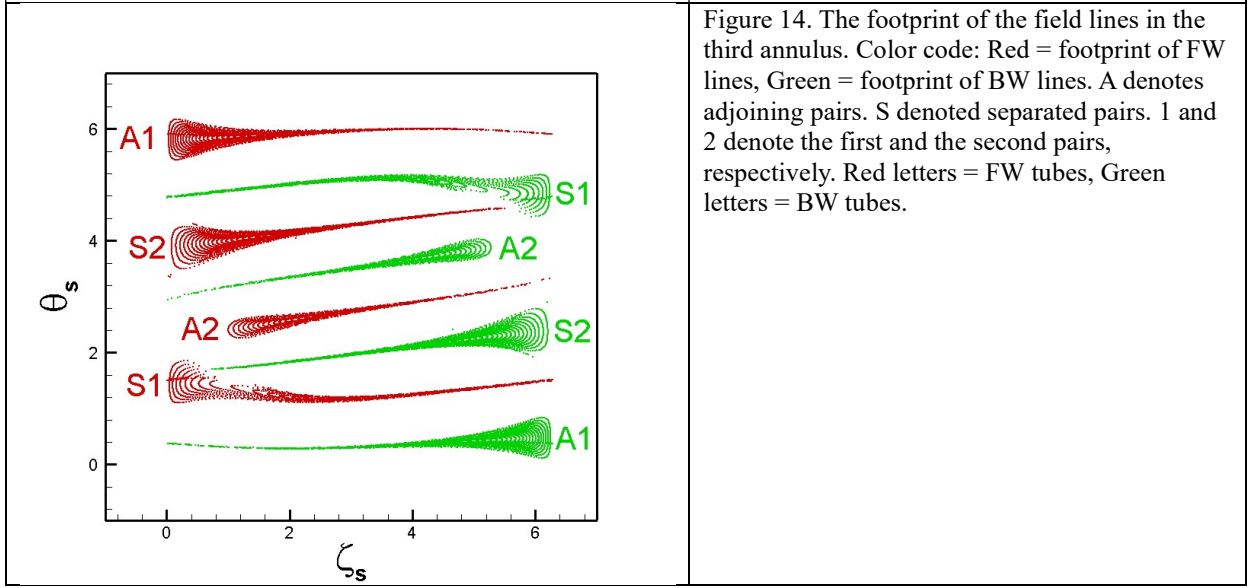

Figure 14. The footprint of the field lines in the third annulus. Color code: Red = footprint of FW lines, Green = footprint of BW lines. A denotes adjoining pairs. S denoted separated pairs. 1 and 2 denote the first and the second pairs, respectively. Red letters = FW tubes, Green letters = BW tubes.

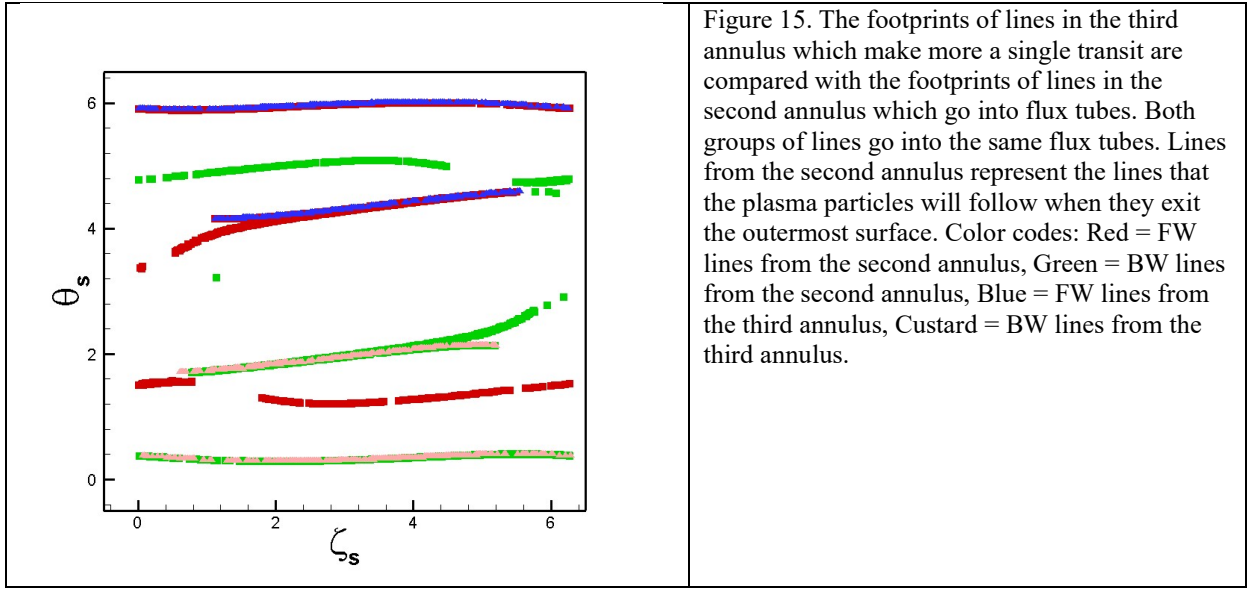

Figure 15. The footprints of lines in the third annulus which make more a single transit are compared with the footprints of lines in the second annulus which go into flux tubes. Both groups of lines go into the same flux tubes. Lines from the second annulus represent the lines that the plasma particles will follow when they exit the outermost surface. Color codes: Red = FW lines from the second annulus, Green = BW lines from the second annulus, Blue = FW lines from the third annulus, Custard = BW lines from the third annulus.

## VI. Conclusions

The important new findings on the nonresonant stellarator divertors are:

First: The outermost perfectly confining surface has the radius of $r/b = 0.81$. The outermost perfectly confining surface has 0.86 of the toroidal flux of the outermost surface.

Second: The outermost perfectly confining surface is surrounded by a layer of sticky and leaky surfaces. This thickness of this layer is about $w/b \cong 0.06$. This layer is bound on the inside by the outermost perfectly confining surface $r/b = 0.81$ and on the outside by the outermost surface $r/b = 0.87$. Toroidal flux of this layer is about 0.14 of the outermost surface. Deep inside this layer, field lines can remain for hundreds of thousands of transits. The number of transits is so great that this layer probably acts as if confining surfaces exist. Segments on these surfaces where there are sharp edges or changes in surface curvatures are the most-sticky. A small fraction (less than about 5%) of field lines in this layer can leak through surfaces and go into magnetic flux tubes and reach the wall.

Third: Outside this leaky and sticky layer, there is a thin layer of thickness $\Delta r/b \cong 0.01$ where the plasma exits the outermost surface of radius $r/b = 0.87$. Toroidal flux of this layer is 0.02 of the outermost confining surface. A large fraction of field lines $\cong 65\%$ in this layer make about 26,000 transits and go into magnetic flux tubes and reach the wall. These tubes form two pairs of flux tubes – one adjoining pair and the other separated pair. Each pair has an outgoing tube and an incoming tube. Magnetic flux in the outgoing tube and the incoming tube of a pair is equal. There is also one pseudo-pair which extrudes a small distance from the outermost surface but does not reach the wall. This was the old picture of nonresonant divertor [25]. The lines from the layer of leaky and sticky surfaces and from the layer of the exiting plasma go into the same flux tubes.

Fourth: Lines in the volume outside these two layers, $r/b = 0.88$ to the wall $r_{WALL}/b = 4$, generate two new pairs of flux tubes. Lines from this large volume go into in all four pairs of flux tubes. Two of these pairs are the same as the old pairs [25]. One of two new pairs is an adjoining pair and the other is a separated. One of the two new pairs is where the old pseudo-pair was. So, the new picture is now that there are in all four pairs of tubes – two are adjoining pairs and two are separated pairs. In this large volume, about 85% of the lines strike the wall before completing even a single transit. The loss of plasma from this wide region is so fast compared to cross-field plasma diffusion that the lines in this region are probably irrelevant to the study of divertors. Toroidal flux of this large volume outside the shifted outermost surface to the wall is about 20 times that of the outermost surface.

Chaos essentially pays no role in the behavior of magnetic field lines in an island divertor. The W7-X has an island divertor. On the other hand, in a nonresonant divertor, the sharp edges on the magnetic surfaces in the edge naturally give rise to chaos in the trajectories of magnetic field lines. In this case, the confinement of plasma by magnetic surfaces is limited by cantori. Cantori are smooth toroidal surfaces that resemble an irrational magnetic surface but with pairs of small slit-shaped holes. These holes are called turnstiles. Equal amounts of magnetic flux can cross these holes going inward or outward. These flux tubes of magnetic field lines leaking through the holes/slits are compact and always come in pairs to preserve the flux, one tube takes the flux outwards and the other tube takes the flux inwards, forming a turnstile. The smallness of the flux tubes collimates the field lines.

In 2018, an efficient mathematical method was developed to simulate the breakdown of magnetic surfaces into regions of chaotic magnetic field lines bounded by cantori penetrated by turnstiles [23]. In this 2018 paper, Boozer and Punjabi derived a simple analytic model for the Hamiltonian for the trajectories of magnetic field lines in nonresonant divertor, Eqn. (1). Punjabi and Boozer calculated pictures of the magnetic flux tubes in nonresonant divertor in [24, 25]. It was found that the there are two true turnstiles, one adjoining turnstile and one separated turnstile; and one pseudo- turnstile in nonresonant divertor [24, 25]. In the 2022 Punjabi and Boozer paper, pictures of the tubes were produced, see Figures 2 (a)-(d) in Ref. [25]. The magnetic footprint is where the compact tubes of collimated field lines intercept the wall. The field lines in the tube strike the wall in helical stripes covering an area on the wall which is an order of 1% of the wall area. The work here shows that there not two but four turnstiles.

Boozer has suggested [28] that the magnetic field line chaos implies that the electric potential required for quasi-neutrality results in Bohm-like diffusion of the plasma with diffusion coefficient $D_q$ determined by electron temperature $T_e$ alone. For chaotic trajectories, neighboring magnetic field lines quickly separate with distance $\ell$ along the line. This means that the electric field across chaotic field lines $\vec{E}_\perp$ becomes arbitrarily large. A balance of terms shows that the $\vec{E} \times \vec{B}$ drift leads to a Bohm-like transport. Inside the chaotic region, this cross-field diffusion is of central importance. The plasma apparently relaxes diffusively between the innermost and the outermost cantori. The effects of diffusion can be studied using the model magnetic field used here together with a Monte Carlo representation of the diffusion.

Although the field lines that take less than a single transit to strike are technically chaotic, meaning exponentially separating with transits, do not have sufficient time to separate. These are 85% of the lines in the third region. The remaining 15% of lines in region 3 and the lines from region 1 and 2 that go into tubes are important for contribution to the Bohm-like diffusion. These groups of lines from the model Hamiltonian and Monte Carlo method for diffusion can give us the plasma transport and plasma footprint in nonresonant divertor.

**Acknowledgements**

This work was supported by US Department of Energy Office of Science grant numbers DE-SC0023548 to Hampton University, DE-SC0024548 to Princeton University, and DE-AC02-09CH11466 to Columbia University. This research used resources of the NERSC, supported by the Office of Science, US DOE, under Contract No. DE-AC02-05CH11231.

**Data availability**

Codes will be available on reasonable requests.

**Conflict of interest**

Authors have no conflict of interest.

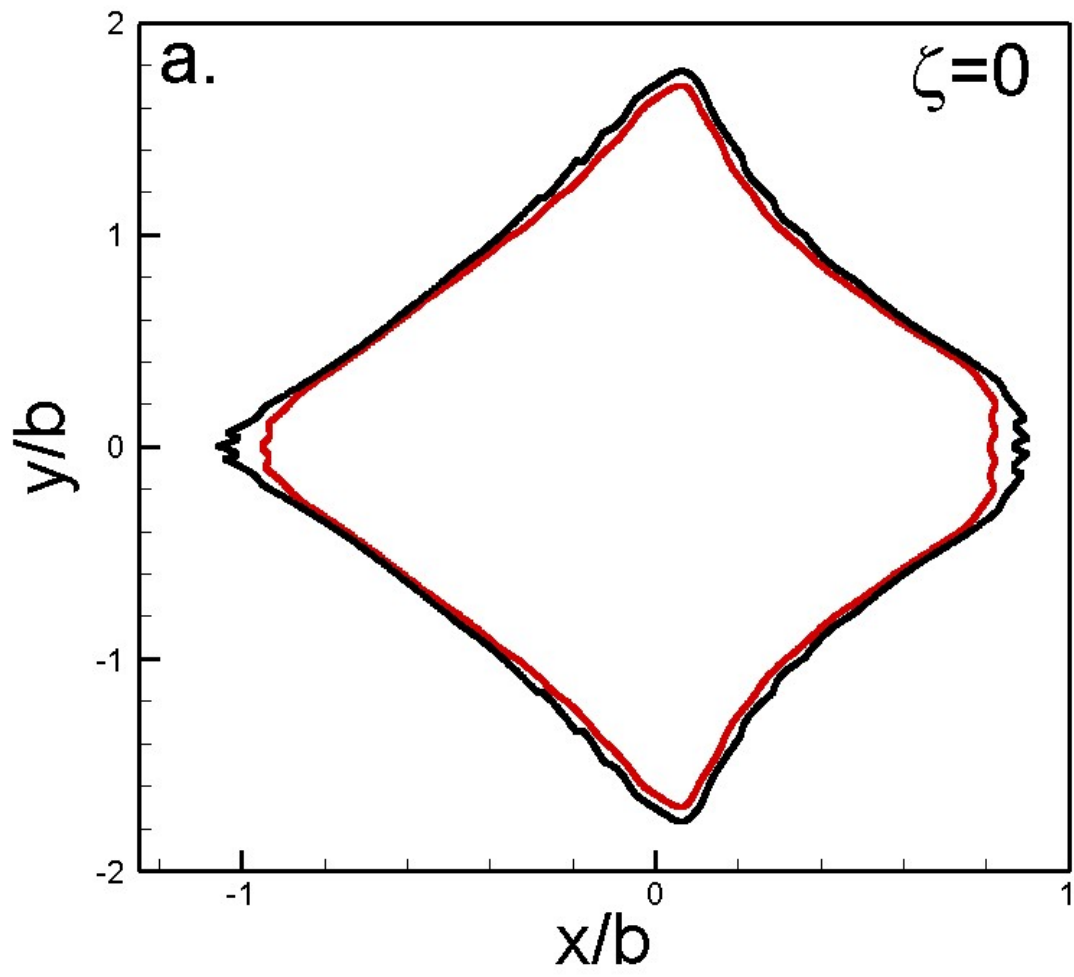

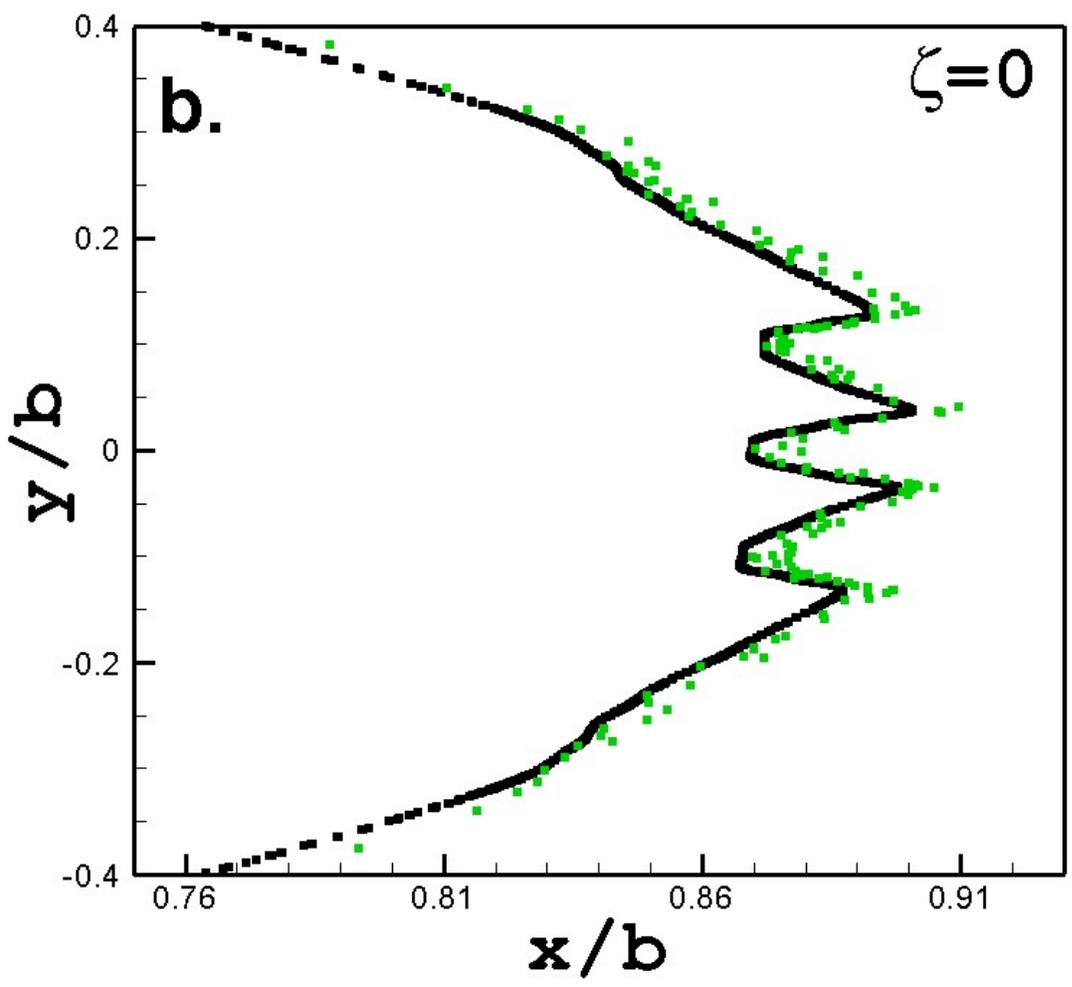

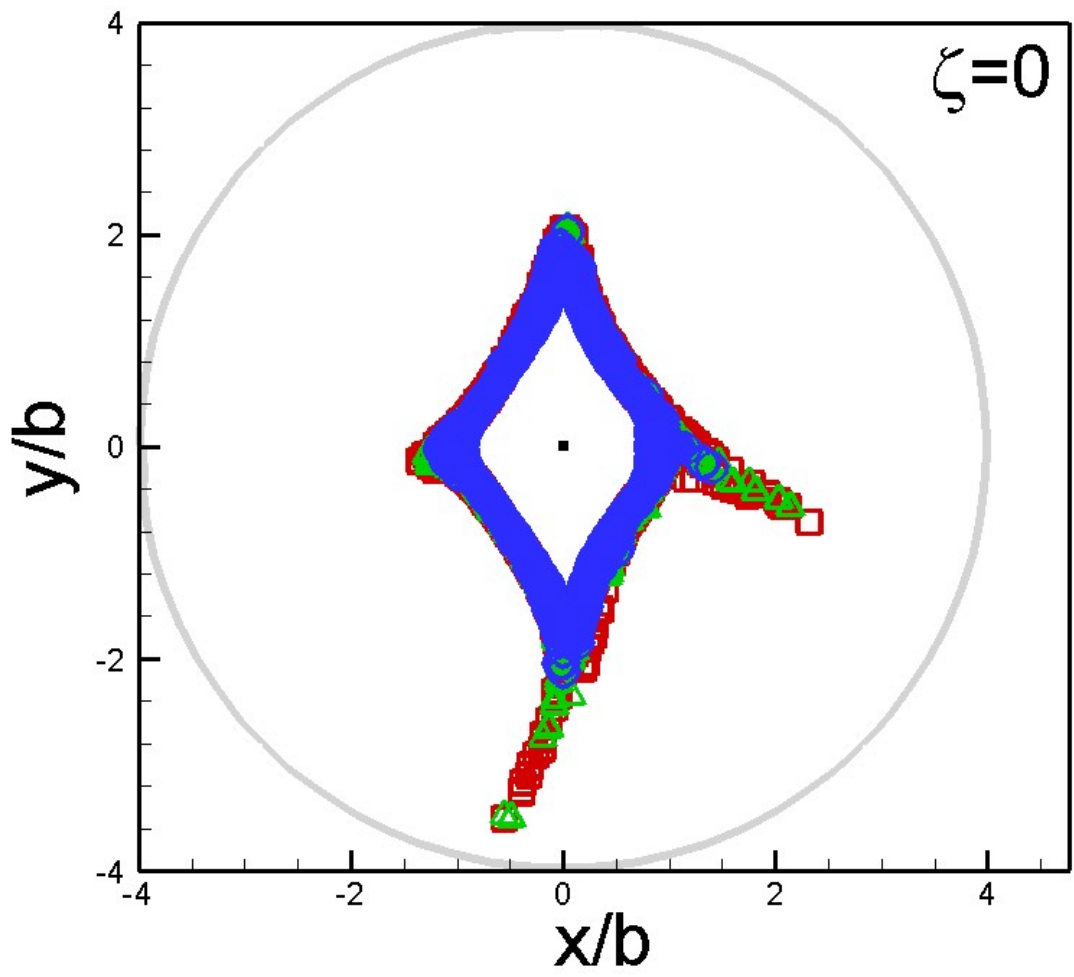

$\tau_{leak} = \zeta_{strike} / 2\pi$

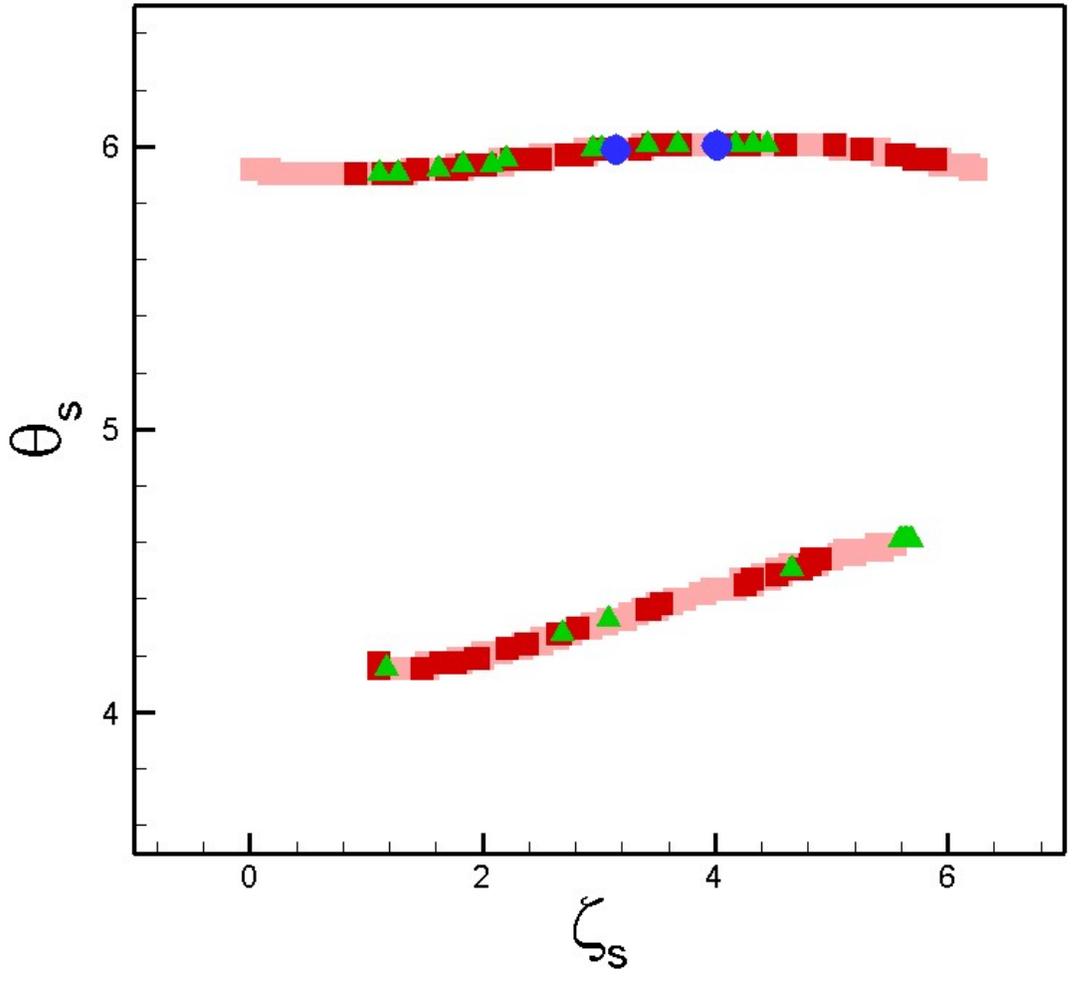

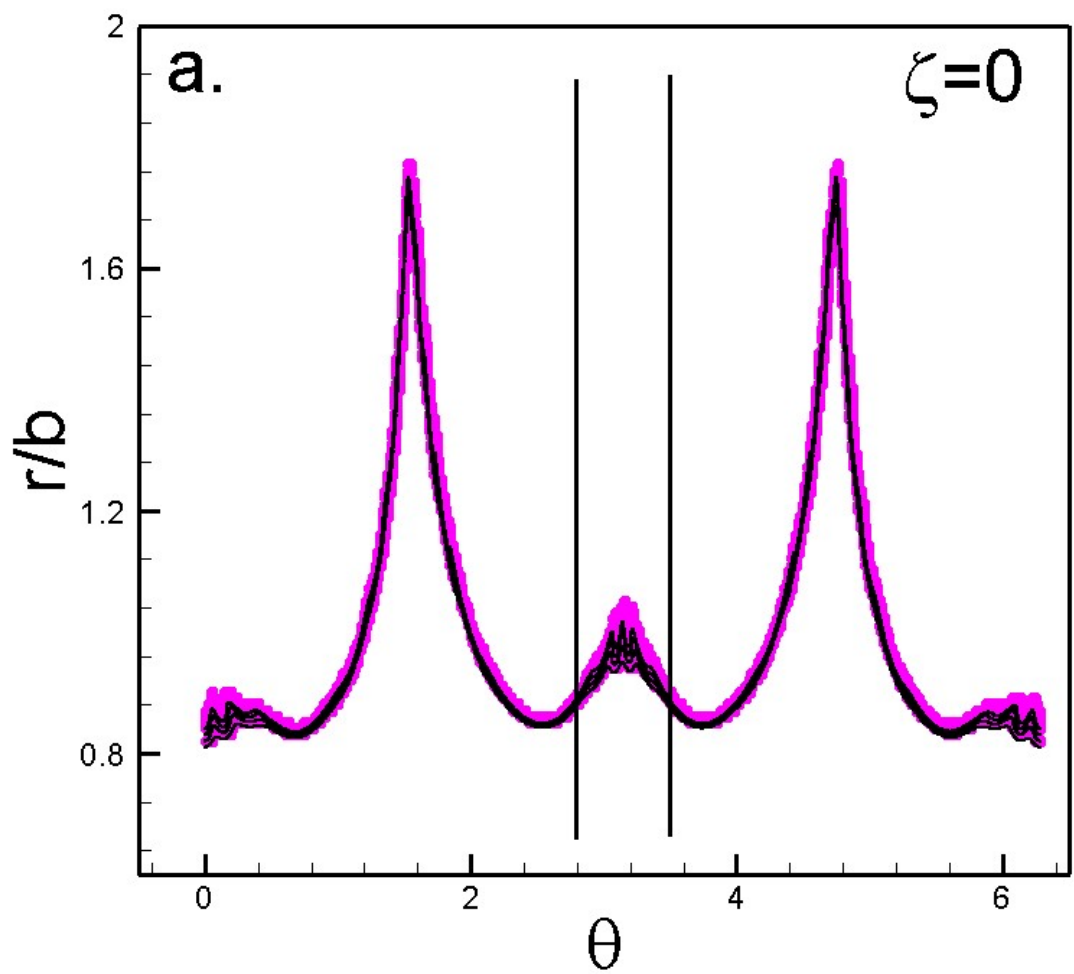

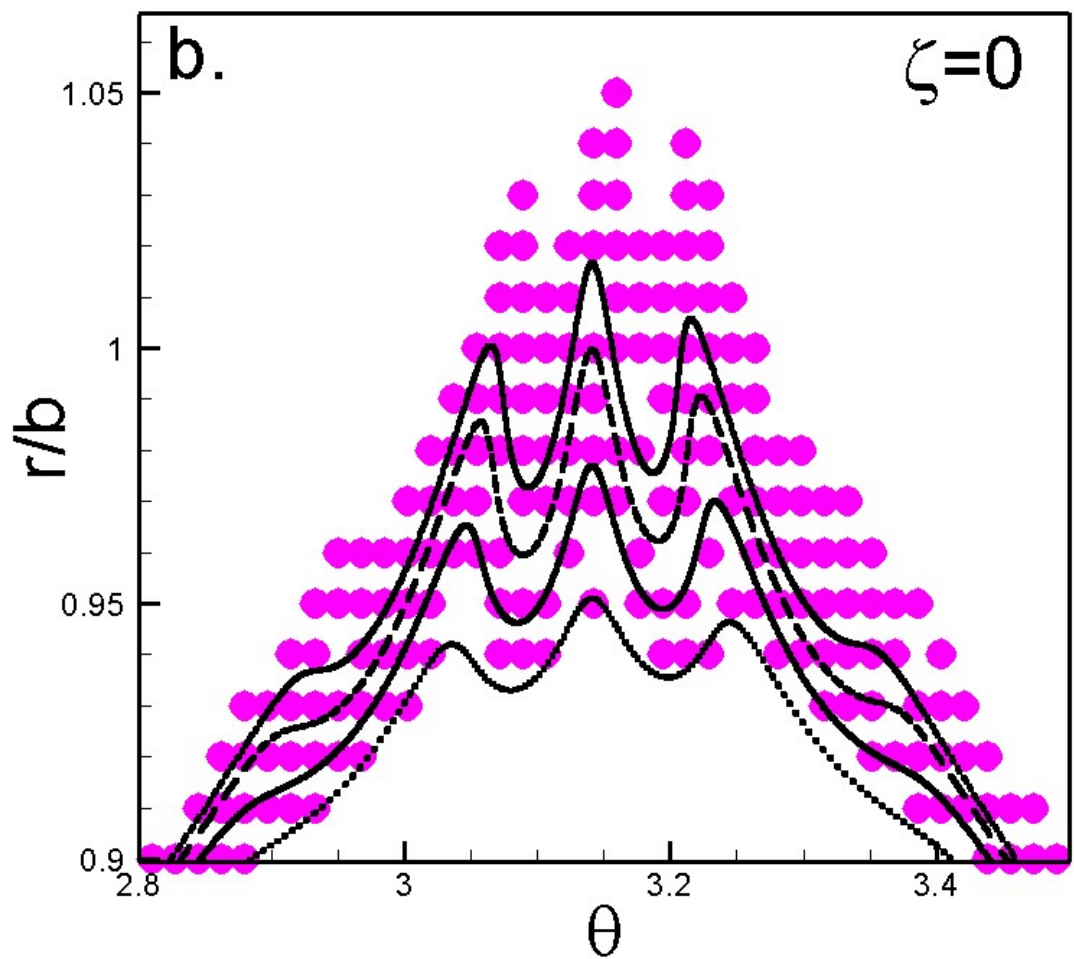

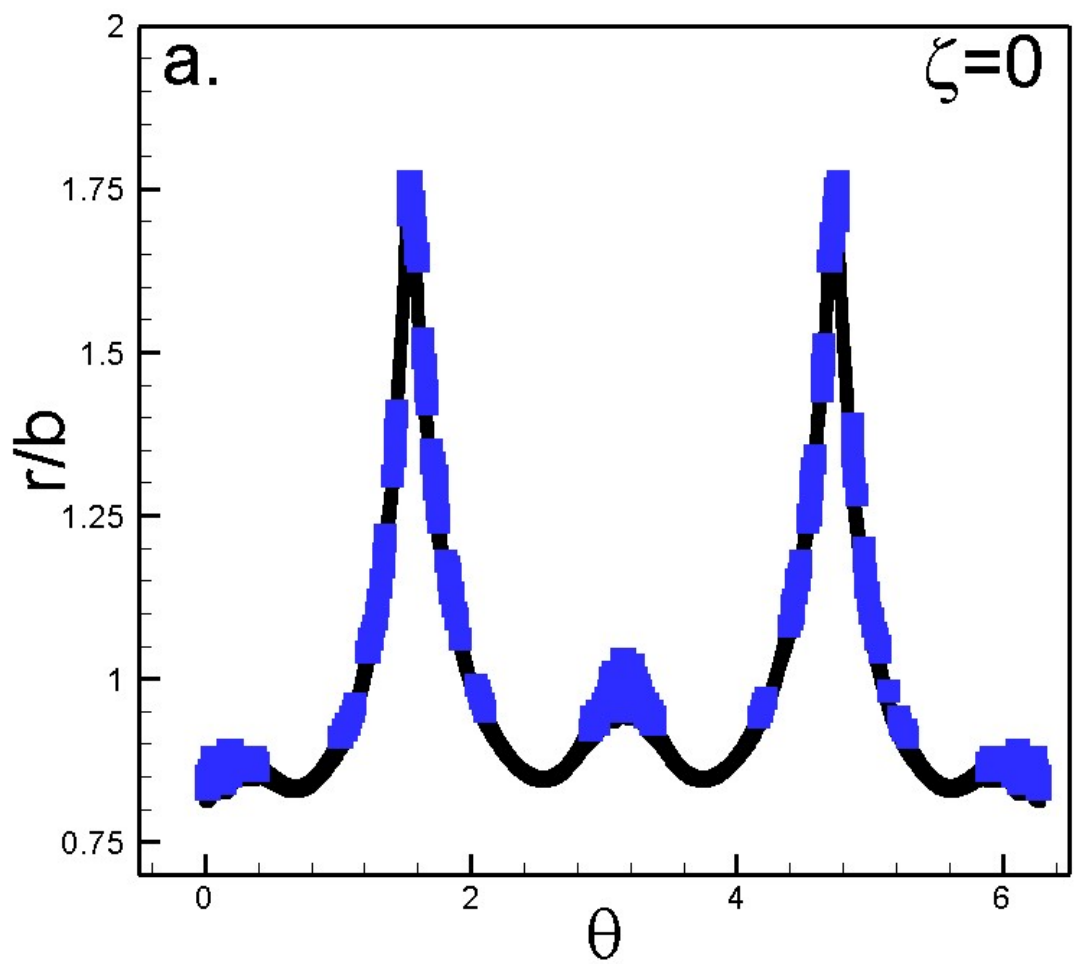

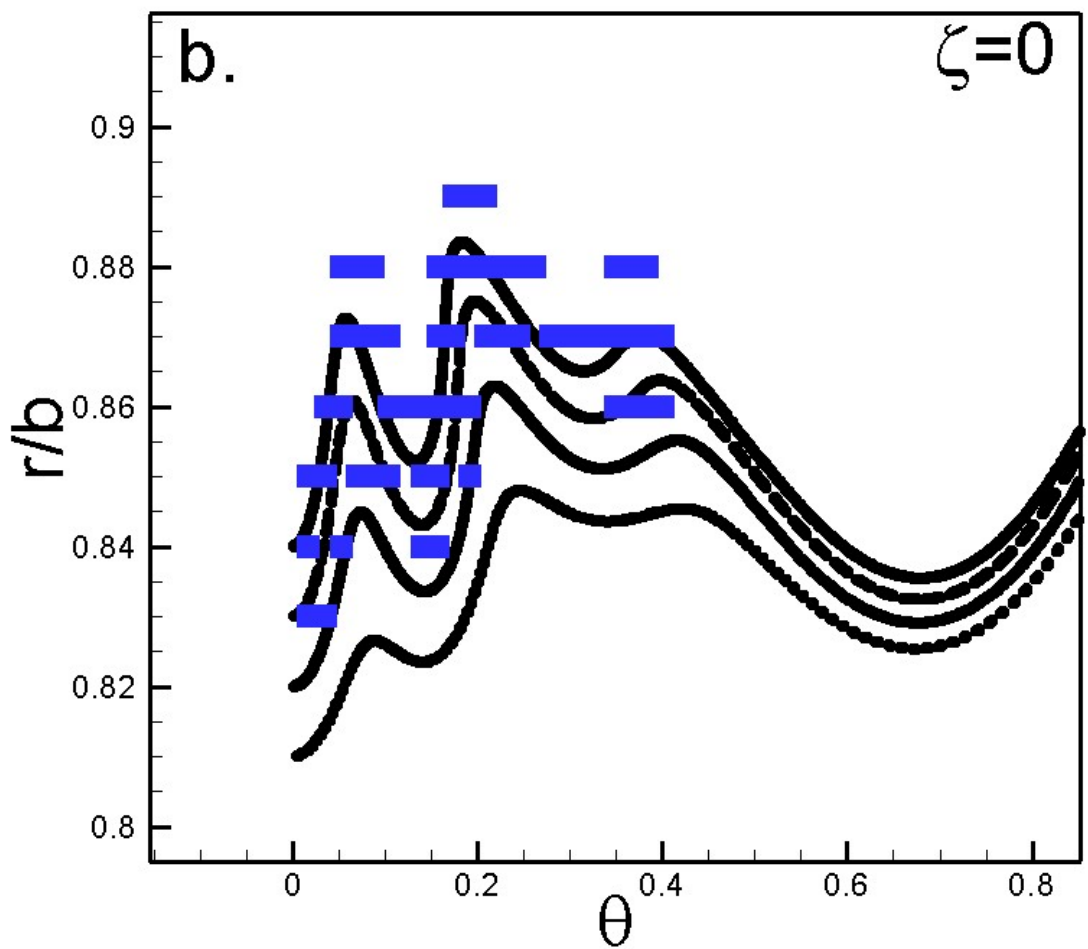

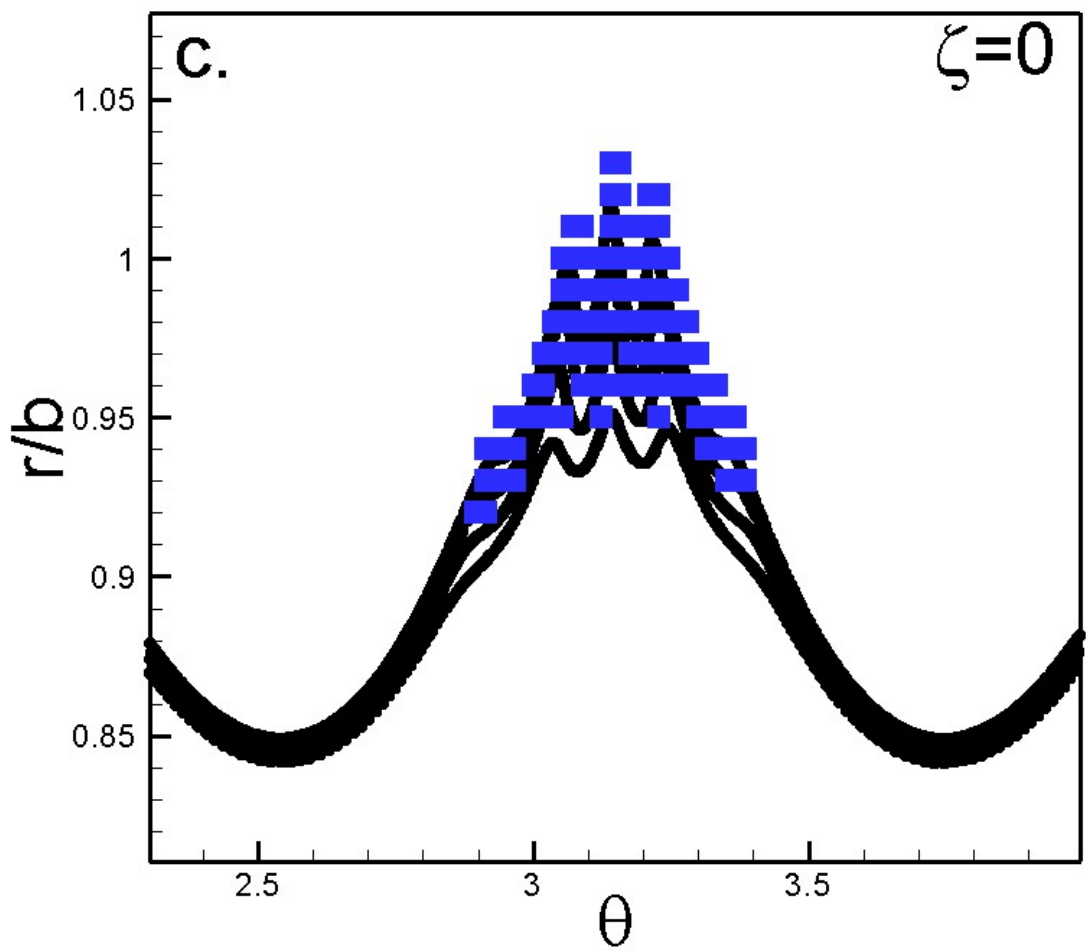

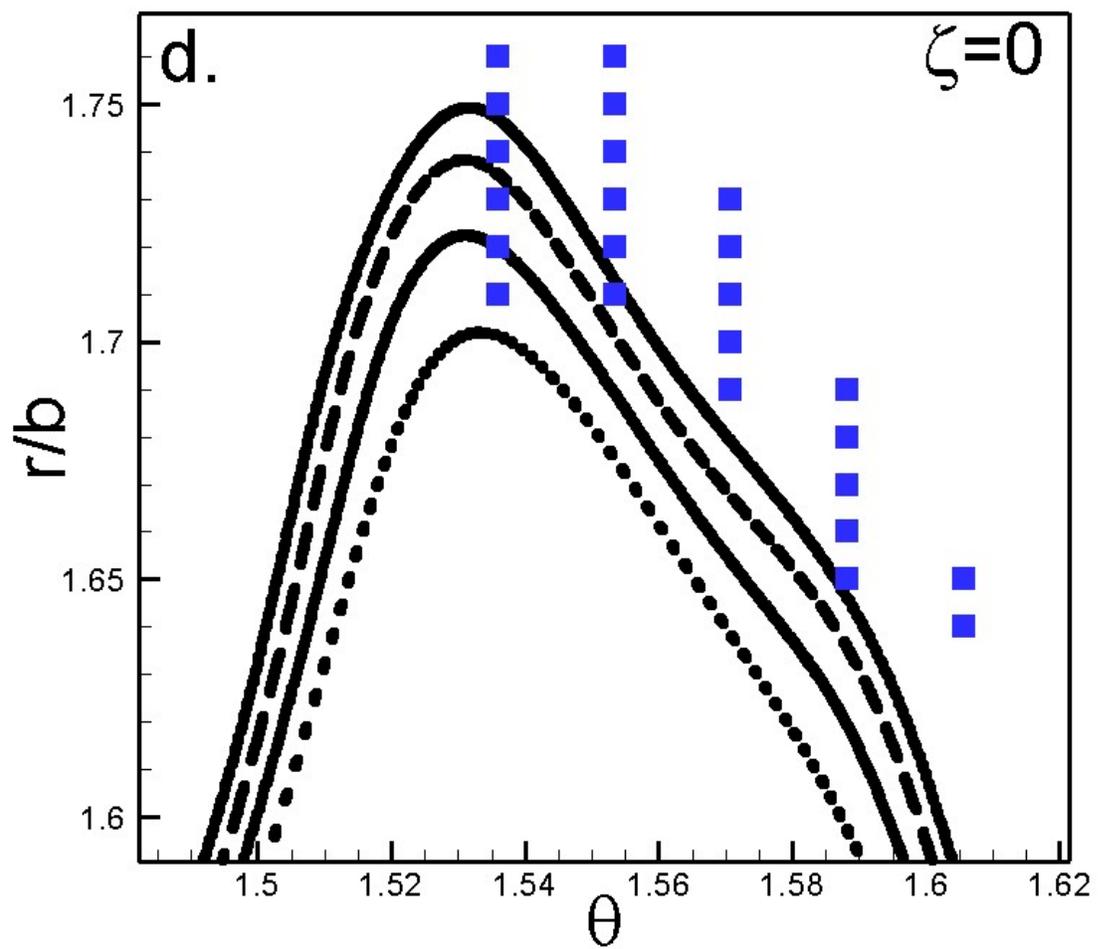

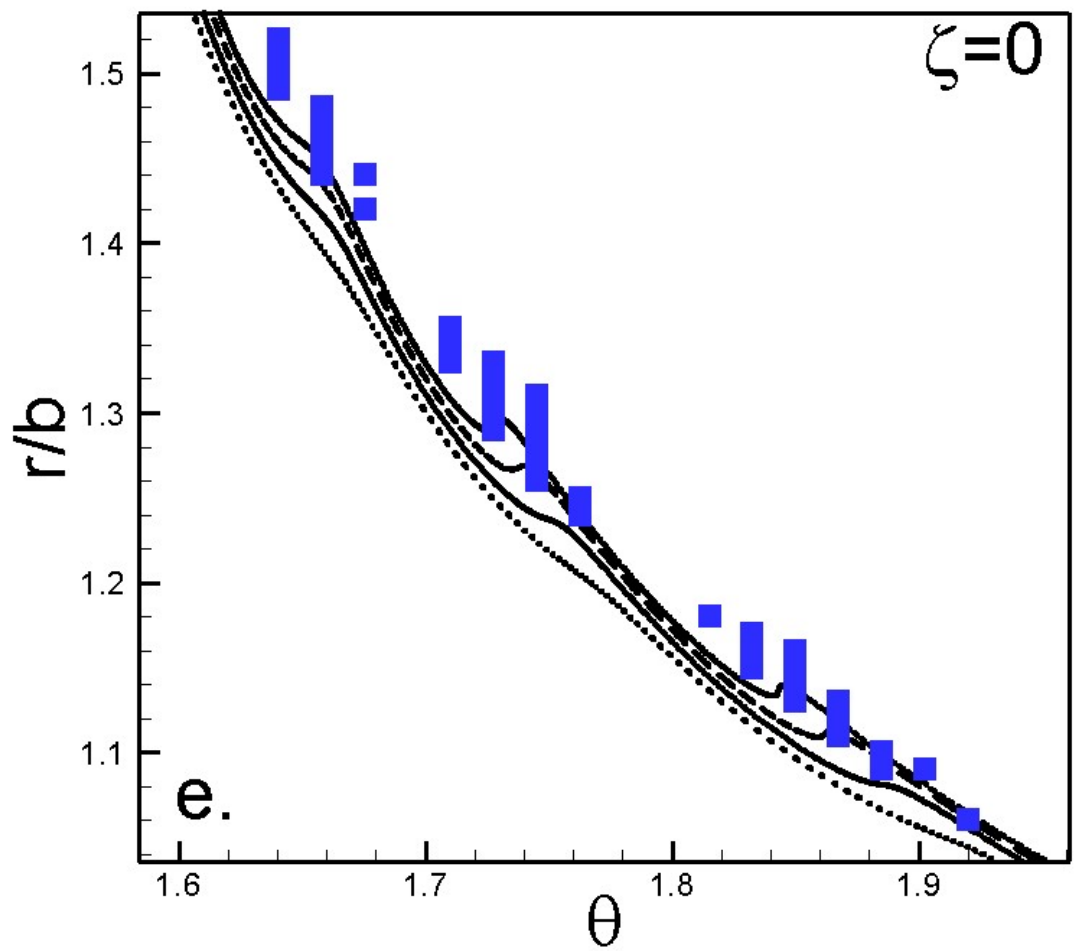

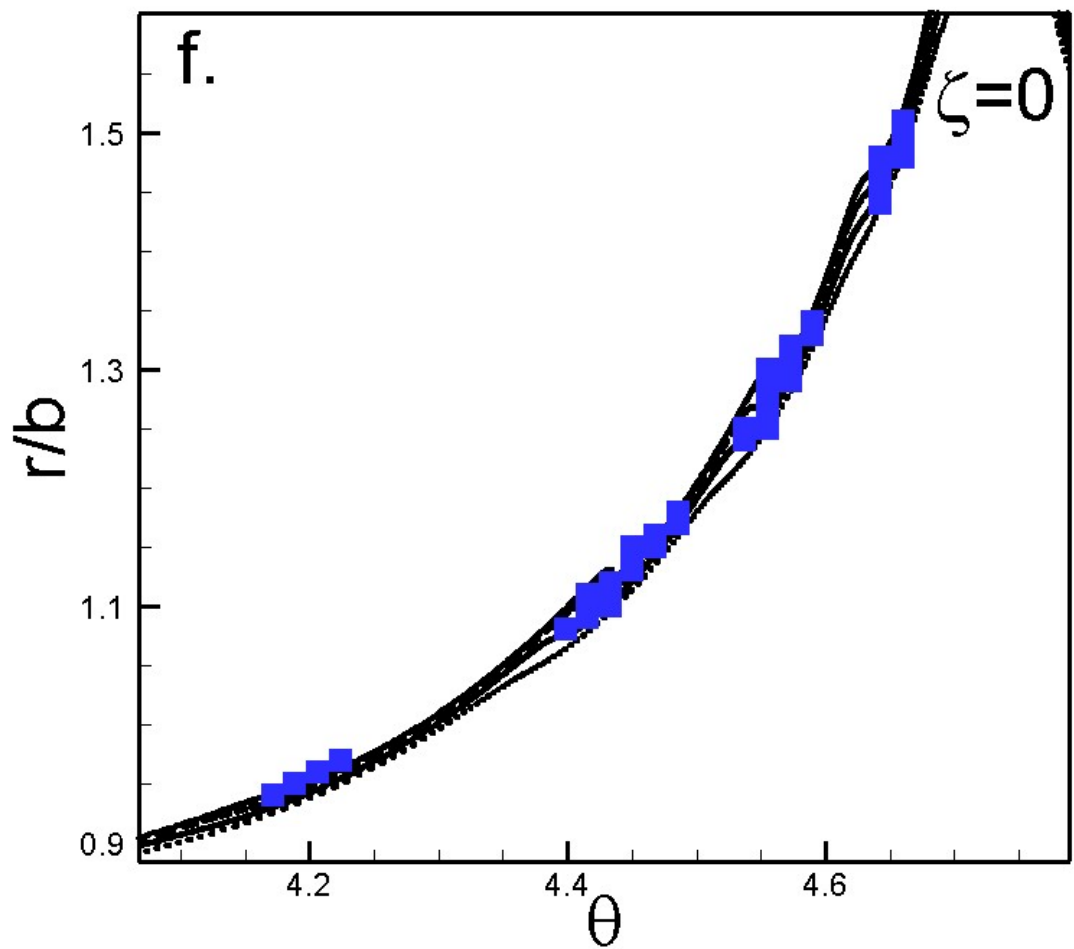

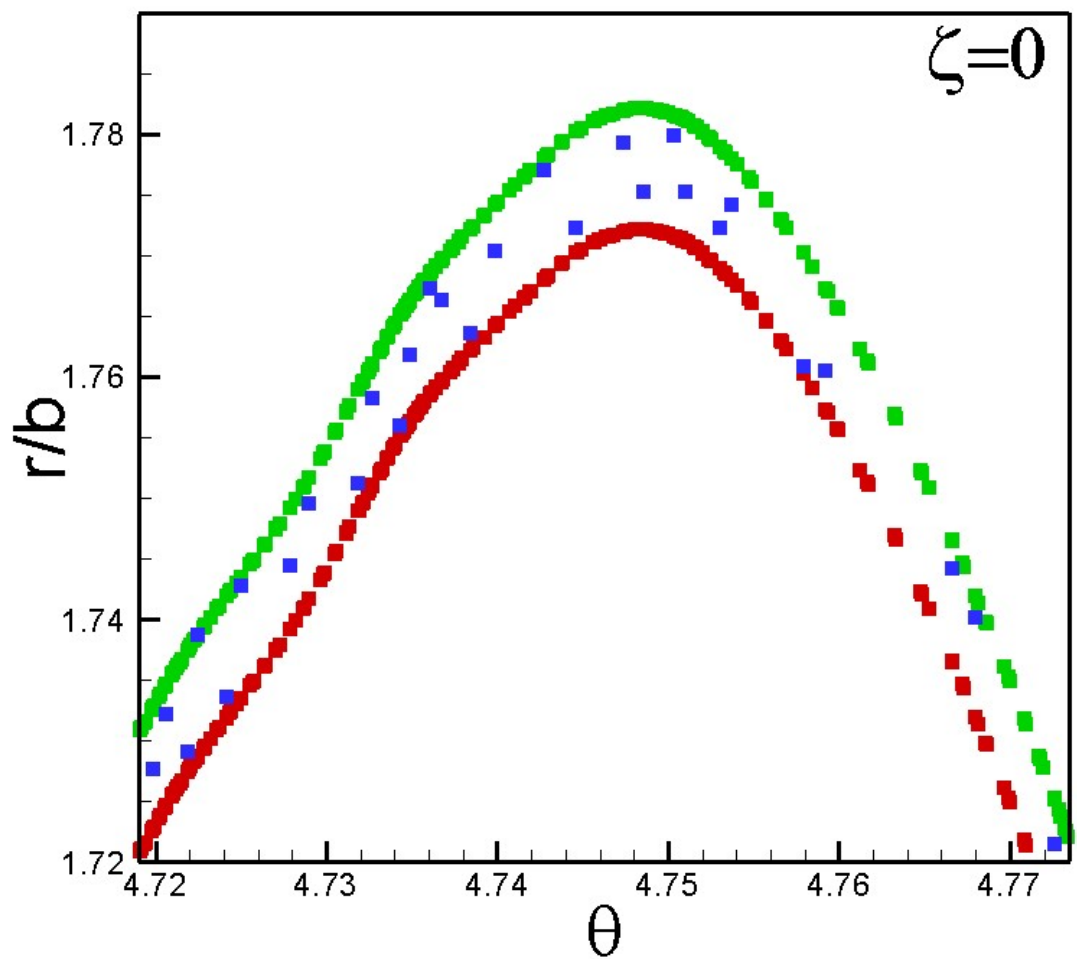

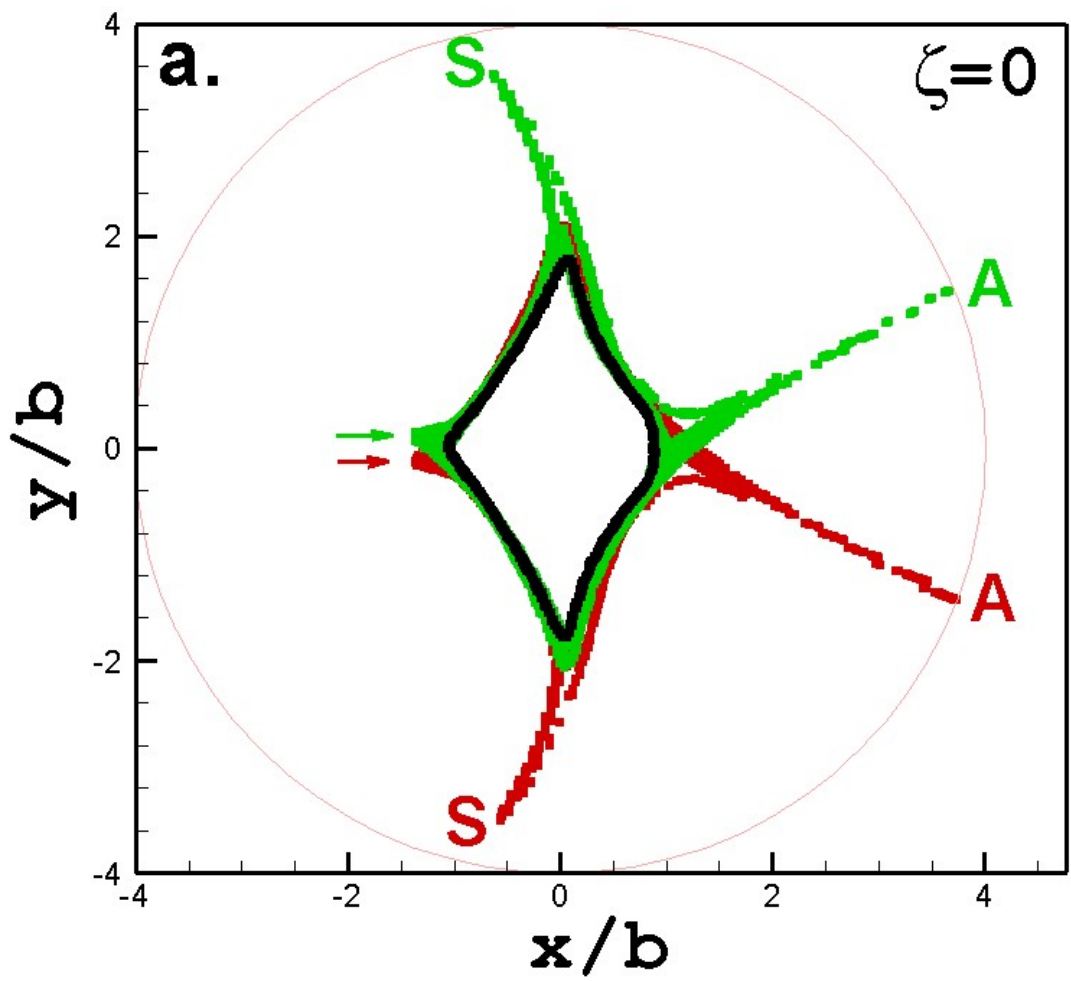

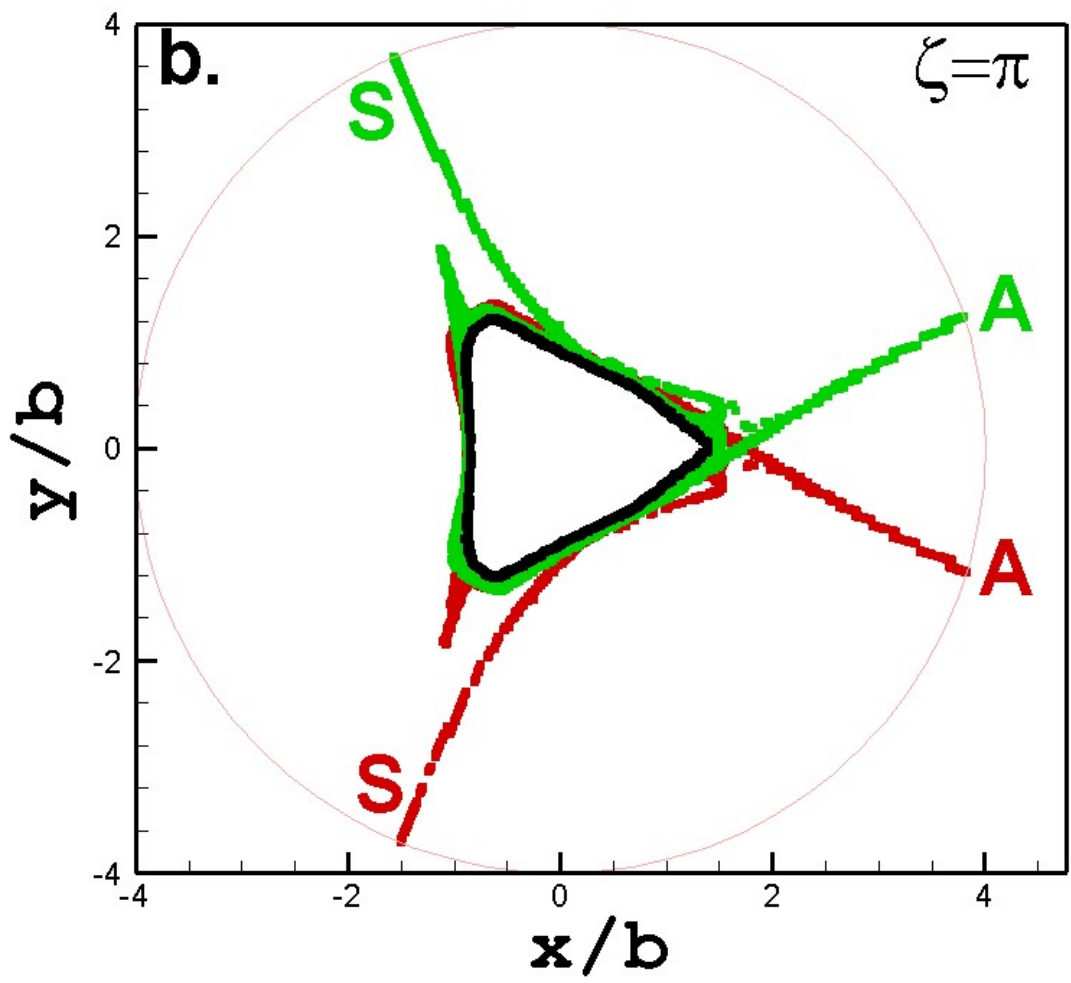

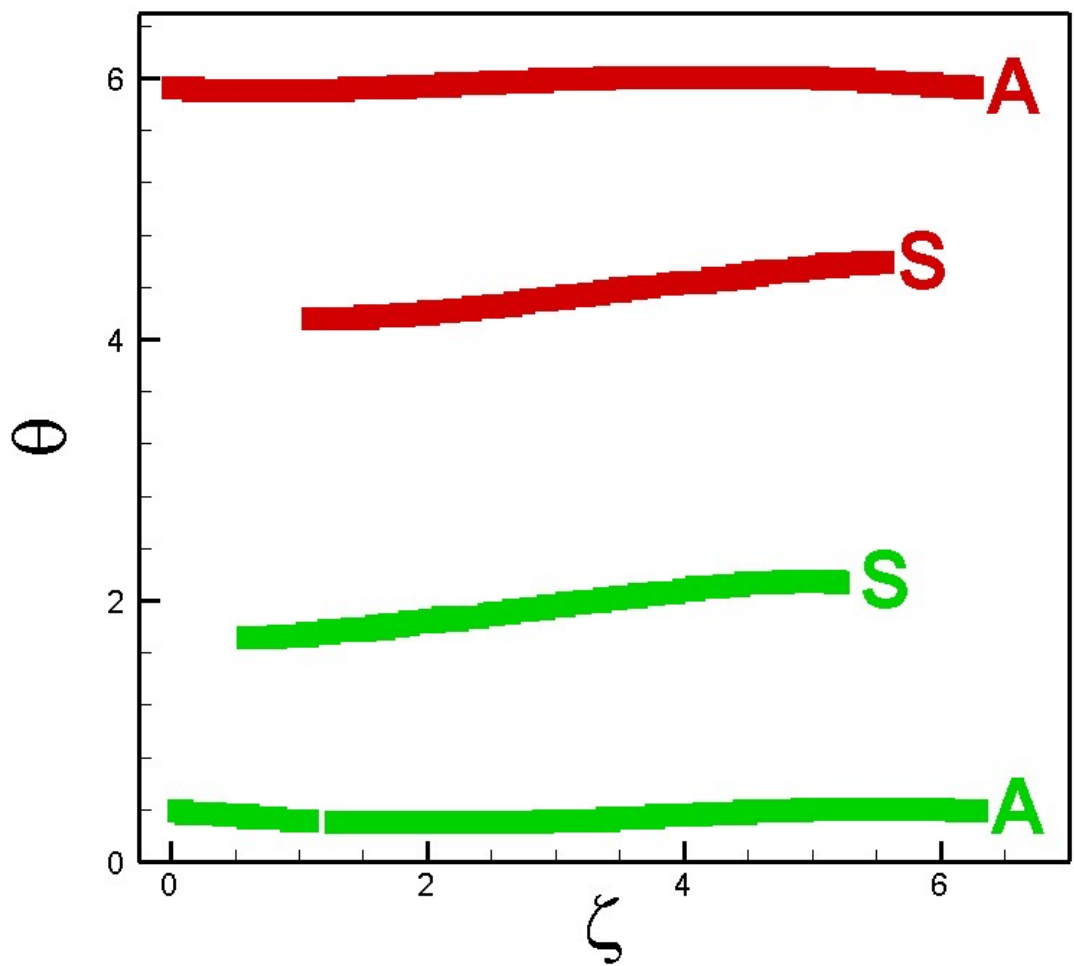

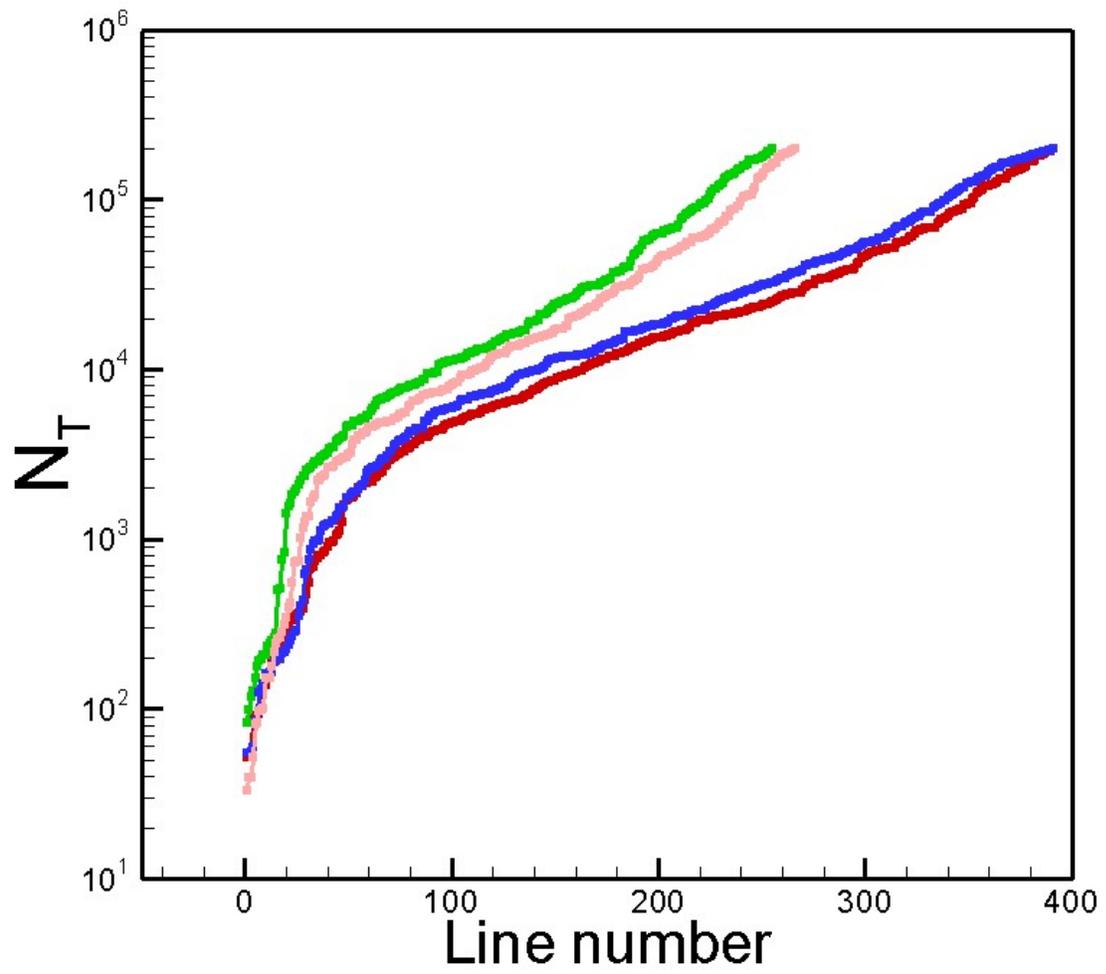

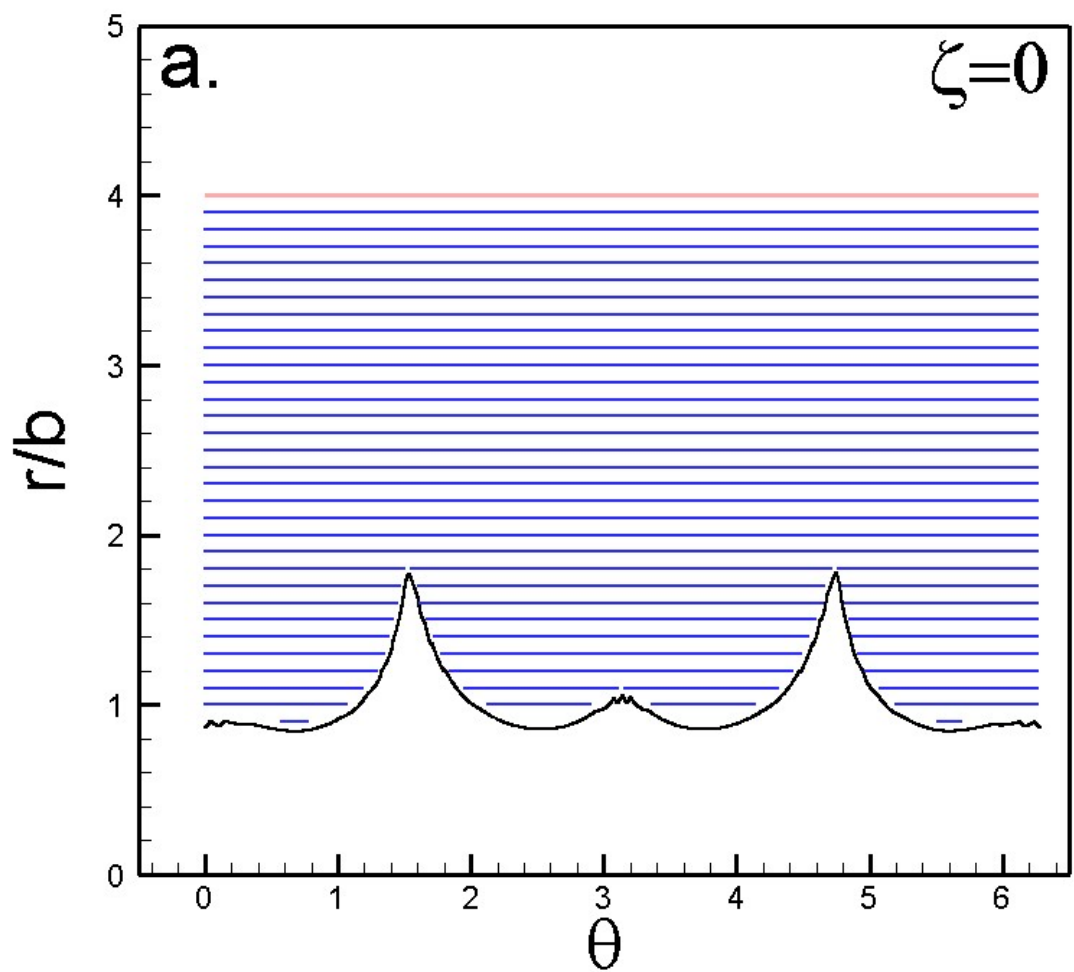

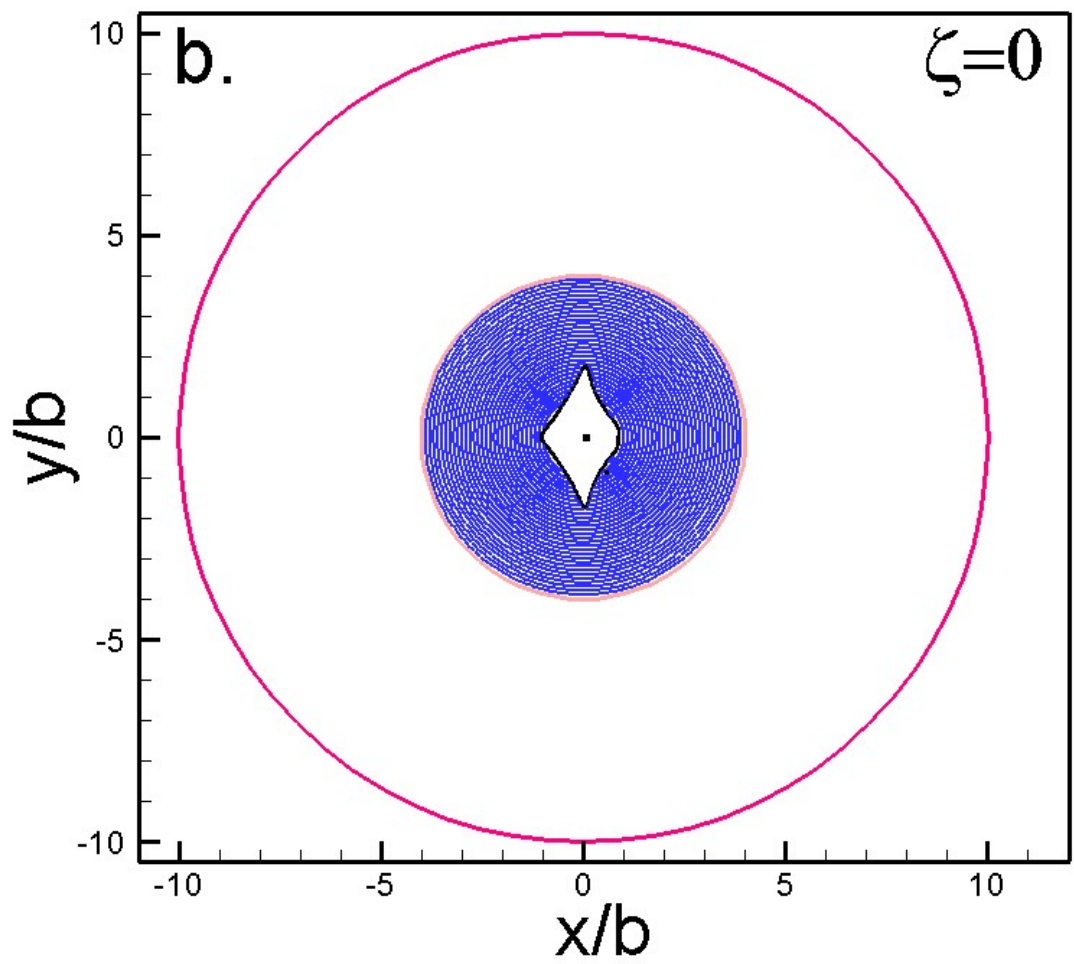

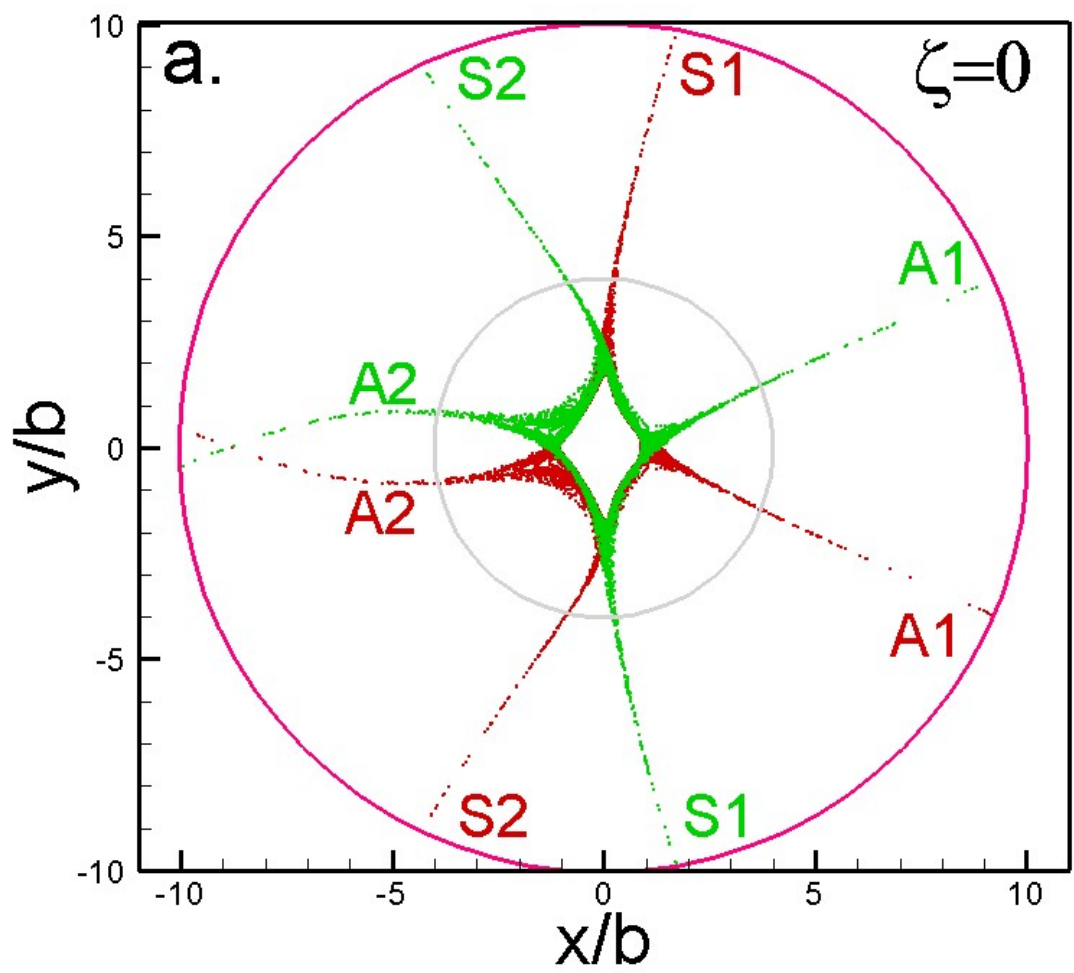

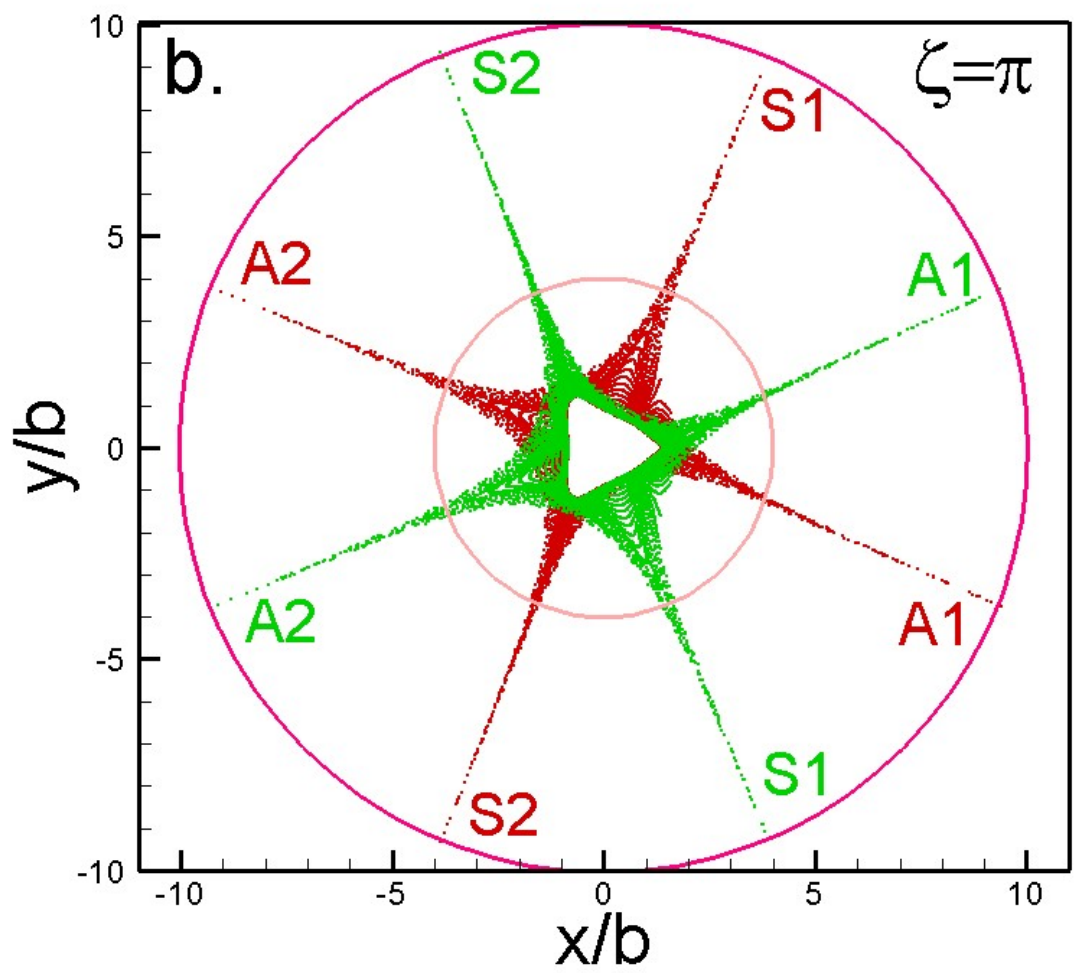

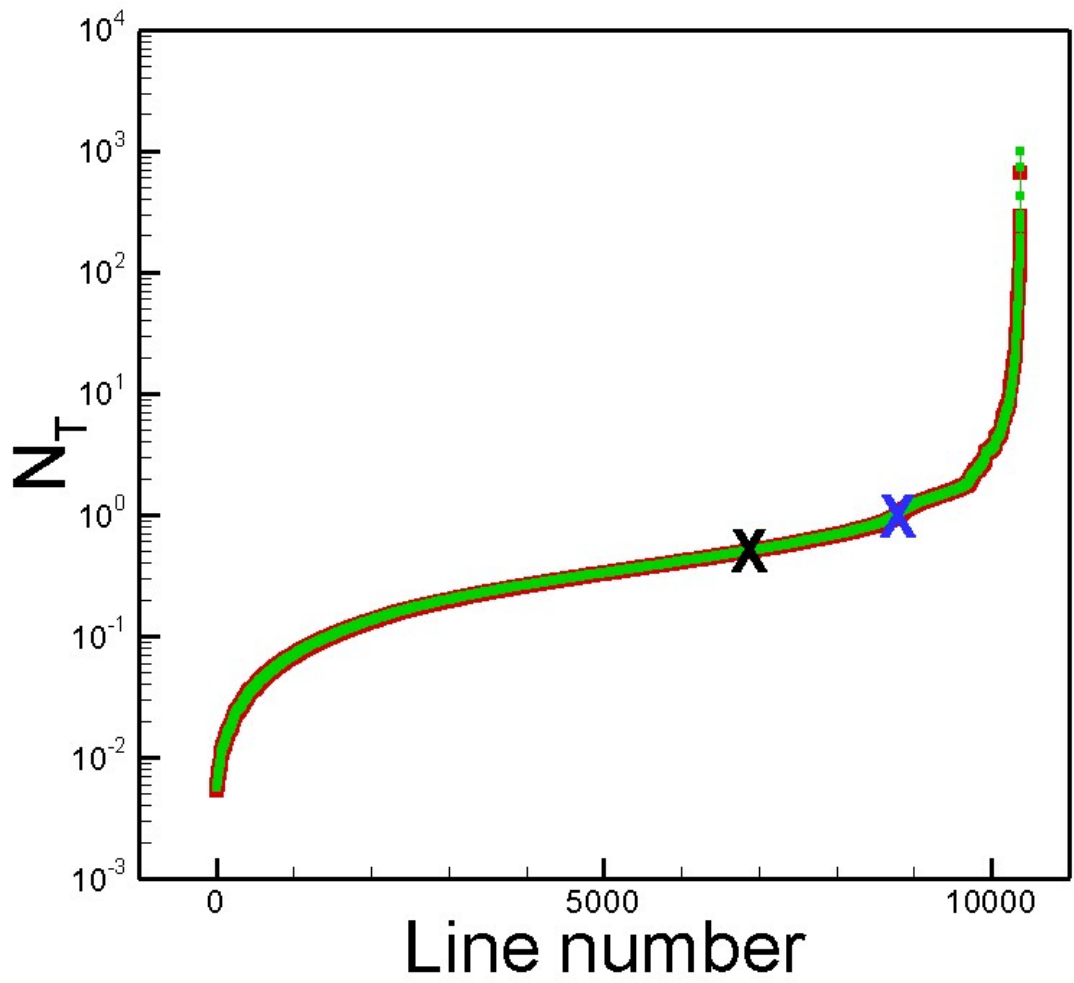

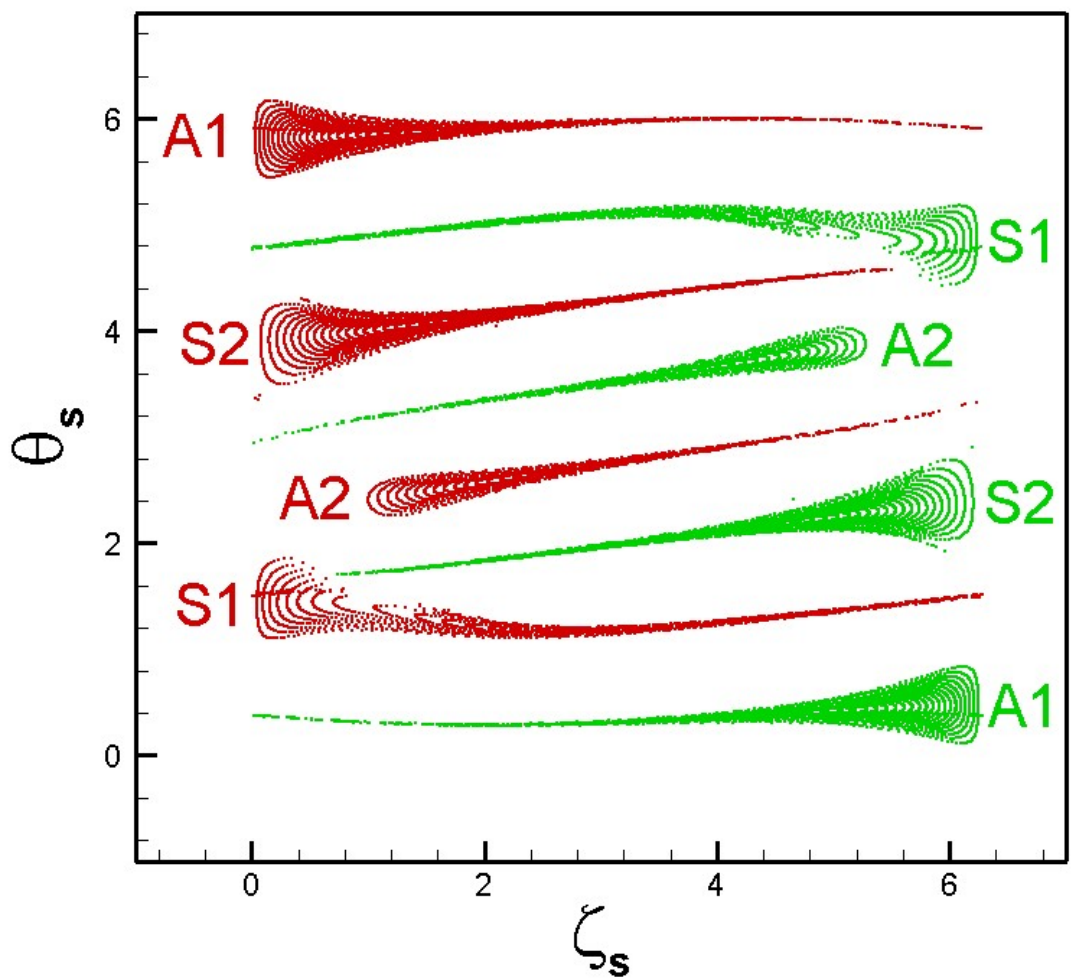

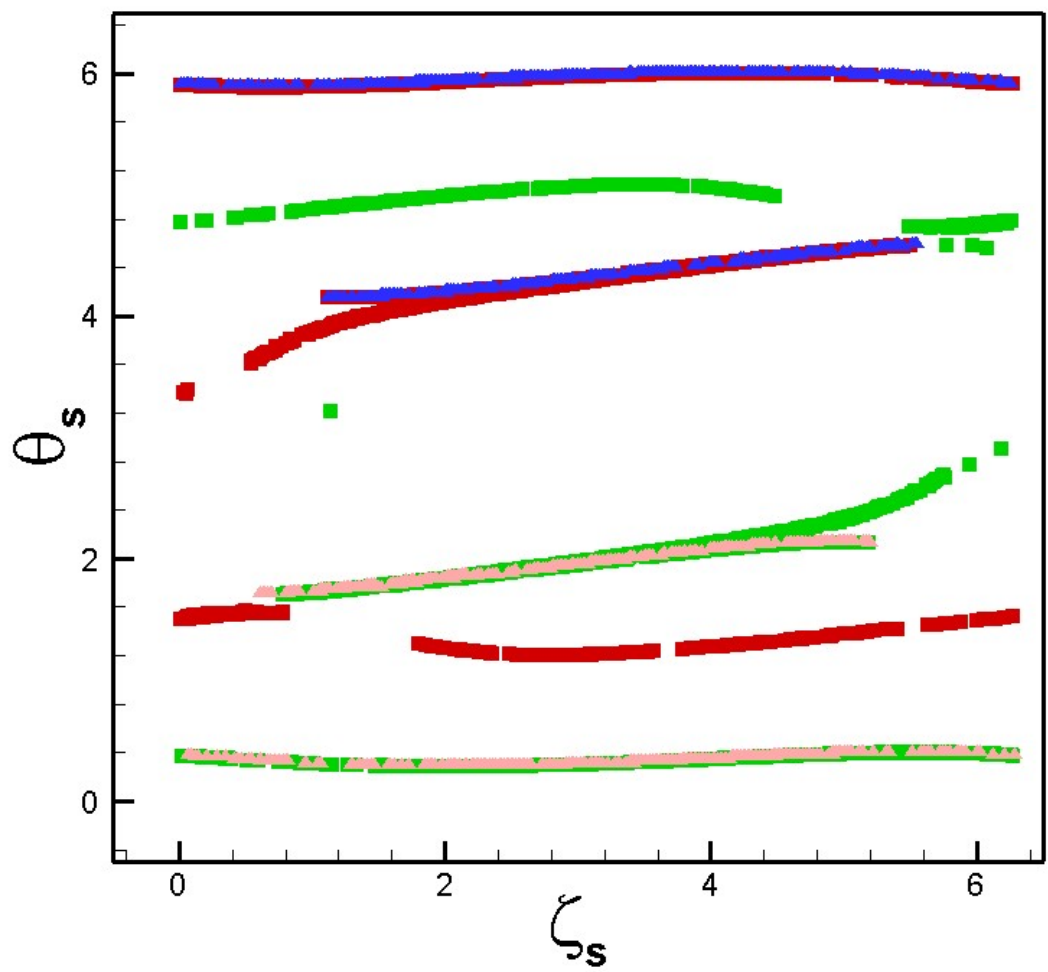